\documentclass[]{aastex701}
\shorttitle{Spatially Resolved Observations of Betelgeuse}
\shortauthors{Matthews et al.}
\submitjournal{AJ}

\usepackage{xcolor}
\newcommand{\lsim}{~\rlap{$<$}{\lower 1.0ex\hbox{$\sim$}}}
\newcommand{\gsim}{~\rlap{$>$}{\lower 1.0ex\hbox{$\sim$}}}
\newcommand{\as}[2]{$#1''\,\hspace{-1.7mm}.\hspace{.0mm}#2$}

\newcommand{\ad}[2]{$#1^{\circ}\,\hspace{-1.7mm}.\hspace{.0mm}#2$}

\newcommand{\MnI}{\mbox{Mn\,{\sc i}}}

\newcommand\kms{\,${\rm km\>s}^{-1}$\>}

\def\up#1{\leavemode \raise.16ex\hbox{#1}}

\begin{document}

\title{New Spatially Resolved Observations of Betelgeuse from the VLA and ALMA: 
Evidence for Atmospheric Perturbations from a Close Companion}

\author[orcid=0000-0002-3728-8082, gname=L. D., sname=Matthews]{L. D. Matthews}
\affiliation{Massachusetts Institute of Technology Haystack Observatory, 99 Millstone Road, Westford, MA 01886 USA}
\email[show]{lmatthew@mit.edu}

\author[orcid=0000-0002-8985-8489, gname=A. K., sname=Dupree]{A. K. Dupree}
\affiliation{Center for Astrophysics $|$ Harvard \& Smithsonian, 60 Garden Street, Cambridge, MA 02138 USA}
\email{adupree@cfa.harvard.edu}

\author[orcid=0000-0002-3880-2450, gname=A. M. S., sname=Richards]{A. M. S. Richards}
\affiliation{Jodrell Bank Centre for Astrophysics, School of Physics and Astronomy, University of Manchester, Manchester, United Kingdom}
\email{a.m.s.richards@manchester.ac.uk}

\author[orcid=0000-0002-2490-1079, gname=W. R. F., sname=Dent]{W. R. F. Dent}
\affiliation{European Southern Observatory/Atacama Large Millimeter/submillimeter Array, Karl-Schwarzschild-Stra{\ss}e 2, 85748 Garching bei M\"unchen, Germany}
\email{billdent42@gmail.com}

\begin{abstract}

We present new spatially resolved observations of the red supergiant Betelgeuse ($\alpha$ Orionis)
at seven frequencies: 22~GHz ($\lambda$1.36~cm); 44~GHz ($\lambda$6.8~mm); 107~GHz ($\lambda$2.8~mm); 136~GHz
($\lambda$2.2~mm); 224~GHz ($\lambda$1.3~mm); 338~GHz ($\lambda$0.89~mm); and 485~GHz ($\lambda$0.62~mm)
using the Karl G. Jansky Very Large Array and the Atacama Large Millimeter/submillimeter Array. These observations
(spanning 2021--2026) enable a characterization of the  mean brightness
temperature  ($T_{\rm~B}$), shape, and symmetry of the star's extended atmosphere. At 22~GHz and 44~GHz
$T_{\rm~B}$ has increased following the historic lows observed
in 2019, just prior to the Great Dimming of late 2019/early 2020. We find that $T_{\rm B}$
peaks at a larger projected radius ($\gsim2.7~R_{\star}$) compared with measurements prior
to 2016, consistent with a possible evolution in the atmospheric density profile.
In addition we find a systematic decrease in $T_{\rm B}$ from $\approx3500$~K at $\sim2.7R_{\star}$ to
$T_{\rm~B}\approx$2220~K at $\sim1.2R_{\star}$, confirming a previously reported
temperature minimum between the photosphere and chromosphere. 
At all observed frequencies $T_{\rm~B}$ is lower than temperatures inferred from ultraviolet
measurements at comparable radii, indicating the continued presence of cooler plasma ($T\le~T_{\rm~eff}$)
mixed with the warmer chromospheric material.
The new measurements yield a global spectral index $\alpha=1.35\pm$0.03, with no evidence for
a change in $\alpha$ over more than two decades in frequency. Recently,
evidence of a close-in companion to Betelgeuse has been reported, with an orbital separation
$r\sim2.3R_{\star}$. We have identified a correlation between the ellipticity of Betelgeuse, as
obtained from radio measurements, and the companion's orbital phase.
The results are consistent with periodic perturbations of the atmosphere by the companion.

\end{abstract}

%



\section{Introduction}\protect\label{sec:intro}
At a distance of  $d\approx$222~pc \citep{Harper2017}, the semi-regular variable star Betelgeuse ($\alpha$~Orionis; HD~39801) is one of the two nearest red supergiants and one of the
largest angular diameter stars as seen from Earth ($\theta_{\star}$=44.2~mas at 2.2~$\mu$m; \citealt{Dyck92}). 
Betelgeuse was the first star other than the Sun to be directly imaged \citep{Gilliland1996}, and since then,
the ability to obtain spatially resolved, multi-wavelength images of Betelgeuse using ground- and space-based telescopes has provided unique opportunities to study the dynamic atmospheric properties of a massive star during the late stages of its evolution (\citealt{Uiten98}; \citealt{Lim1998}; \citealt{Lobel2000}; \citealt{Ohnaka2009}; \citealt{Dupree2020}; \citealt{Montarges2021}; \citealt{Matthews2022}). 
Here we build upon these efforts by presenting new, spatially resolved observations at centimeter (cm), millimeter (mm), and submillimeter (sub-mm) wavelengths.\footnote{For brevity we hereafter collectively refer to these as ``radio'' wavelengths.} 

Radio continuum observations provide a unique and powerful diagnostic probe of the extended atmospheres of red supergiants. The radio emission arises from heights above the classical photosphere (e.g., \citealt{Alt1979}; \citealt{Skinner1997}; \citealt{Lim1998}), and owing to the wavelength dependence of the opacity in the radio portion of the spectrum, measurements at successively longer wavelengths probe material at increasing stellar radii, from $r\sim1.2R_{\star}$ at sub-mm wavelengths out to several times $R_{\star}$ in the cm bands. This allows spatially resolved, multi-frequency radio measurements to characterize the mean brightness (electron) temperature as a function of radius (e.g., \citealt{Lim1998}; \citealt{OG2015}, \citeyear{OG2020}) and to provide constraints on models of the temperature and density structure of the star's extended atmosphere (\citealt{Skinner1997}; \citealt{Harper2001}; \citealt{Dent2026}).  Because the radio emission corresponds to the Rayleigh-Jeans tail of the Planck function, its brightness is directly proportional to the mean electron temperature, in contrast to the ultraviolet, which preferentially traces hotter and denser material \citep[see e.g.,][]{Linsky2017}. At a given height above the photosphere, radio observations
are thus sensitive to plasma whose mean temperature is cooler than the chromospheric component that gives rise to ultraviolet continuum and emission lines (\citealt{Lim1998}; \citealt{Harper2001}; \citealt{HarperBrown2006}). Finally, spatially resolved imaging with radio interferometers provides a means to search for evidence of asymmetries, hot spots, and other features that have long been predicted to manifest in the atmospheres of red supergiants (e.g., \citealt{Schild1975}; \citealt{Gray2000}; \citealt{Chiavassa2009}; \citealt{Freytag2024}).

The long baseline configurations of the Karl G. Jansky Very Large Array (VLA)\footnote{The VLA of the National Radio Astronomy
  Observatory (NRAO) is operated by Associated
  Universities, Inc. under cooperative agreement with the National
  Science Foundation.} and the Atacama Large Millimeter/submillimeter Array (ALMA) both provide sufficient angular resolution to spatially resolve the atmosphere of Betelgeuse at multiple wavelengths, as demonstrated in previous studies of the continuum emission from this star (\citealt{Lim1998}; \citealt{OG2015}, \citeyear{OG2017}; \citealt{Matthews2022}). However, 
for pulsating variable stars like Betelgeuse, multi-epoch observations are of considerable interest to document variations that may occur over the course of the stellar pulsation cycle and/or as a result of secular changes---which may appear relatively suddenly, or manifest gradually over years, or even decades. 

During late 2019/early 2020, Betelgeuse underwent its most dramatic optical dimming in more than a century of recorded photometric monitoring, reaching a minimum $V$-band brightness of 1.614$\pm$0.008~mag between 2020 February 7--13 (\citealt{Guinan2020}, \citealt{Dupree2020}) before rebrightening. This ``Great Dimming'' has since been attributed to 
a surface mass ejection produced by the passage of convectively-induced atmospheric shock waves (\citealt{Kravchenko2021}; \citealt{Dupree2022}). The dimming of the star is thought to have resulted from ensuing increases in molecular line opacity and/or cooling of the ejecta, leading to dust formation and increased line-of-sight extinction (\citealt{Harper2020}; \citealt{Montarges2021}; \citealt{Davies2021}; \citealt{Dupree2022}; \citealt{Granzer2022}; \citealt{Taniguchi2022}). 
Although the brightness of Betelgeuse at visible wavelengths subsequently returned to its historic levels, persistent changes in the star's photometric and radial velocity variations have occurred, including a decrease in the amplitude of the photospheric radial velocity variations from $\sim$10~\kms\ to $\sim$5~\kms\ and a changeover of the $\sim$400~day fundamental pulsation mode  \citep{Joyce2020} to pulsations with half that period, i.e., a first overtone mode (\citealt{Dupree2022}; \citealt{Granzer2021}, \citeyear{Granzer2022}; \citealt{MacL2023}; \citealt{Jadlovsky23}, \citeyear{Jadlovsky24}; \citealt{DupreeGal}). This implies that
the events responsible for the Great Dimming have produced 
persistent changes in the atmosphere of the star (see e.g., \citealt{Montarges2021}, \citealt{Dupree2022} for discussion). Consequently there is heightened interest in comparing the recent radio continuum properties of Betelgeuse with those measured prior to the Great Dimming. Such comparisons have the potential to improve our understanding of the longer-term implications of such events on timescales from months to years---both for Betelgeuse and for other mass-losing red supergiants.

In 2019 August, \cite{Matthews2022} found the brightness temperature of Betelgeuse at 44~GHz ($\lambda$7~mm) to be at its lowest measured value in the past 30 years. Furthermore, they found relative brightness temperature at 22 ($\lambda$1.3~cm) and 44~GHz to be atypical of past trends. These VLA measurements provided one of several pieces of evidence that the recent passage of a strong shock wave had produced perturbations of the density and temperature structure of the atmosphere between $r\sim2R_{\star}$ to $3R_{\star}$ just prior to the onset of the Great Dimming (\citealt{Dupree2020}; \citealt{Dupree2022}; \citealt{Kravchenko2021}; \citealt{Matthews2022}).
 These results have motivated us to pursue additional spatially resolved observations of the star in the ultraviolet (\citealt{Dupree2024}, \citeyear{Dupree2026}) and at radio wavelengths with the VLA to determine whether the flux density and brightness temperature have remained low or have since ``recovered'' to values more consistent with trends over the past three decades. We have also expanded our radio studies to higher frequencies with ALMA to enable a more comprehensive characterization of the properties of the extended atmosphere as a function of depth, including the poorly understood region corresponding to the temperature minimum between the chromosphere and the photosphere (see \citealt{Harper2001}; \citealt{OG2017}). 
 
 Recently, attention to Betelgeuse has been further heightened by multiple lines of evidence pointing to the existence of a close-in binary companion. In addition to its 400~day period (see above) Betelgeuse has long been known to exhibit brightness and
radial velocity variability over a timescale of $\sim$2100 days, though
the origin of the long secondary period remained unclear.  However, the analysis
of more than a century of photometric, radial velocity, and
astrometric measurements, coupled with the evaluation of competing theoretical models, led to
two publications that independently concluded that Betelgeuse harbors a close-in companion that can account for its secondary period (\citealt{Goldberg2024}; \citealt{Mac2025}).  This was recently confirmed through direct
detection of the companion at optical wavelengths using adaptive optics (\citealt{Montarges2026}; see also Section~\ref{companion} below).
  As described below, the high quality of our new VLA and ALMA observations allows us to provide additional corroborating evidence for the existence of a close-in binary companion to Betelgeuse.
  
In the present paper we focus on the derivation of basic stellar parameters as a function of time and frequency based on our new VLA and ALMA observations. We compare the results to similar observations obtained prior to the Great Dimming. We  also consider our latest measurements in the context of Betelgeuse's newly discovered companion.
 More detailed analyses of the images and molecular line spectra from the highest frequency ALMA data (234, 338, and 485~GHz) are presented elsewhere (\citealt{Dent2024}, \citeyear{Dent2026}).
 
\section{Observations}\protect\label{sec:observations}
\subsection{VLA}
Observations of Betelgeuse at 44~GHz ($Q$ band; $\lambda\sim$7.5~mm)
and 22~GHz ($K$ band; $\lambda\sim$1.4~cm) were carried out in 2023, 2024, and 2026, respectively using the VLA in its most extended (A) configuration. Antenna separations ranged from
0.68 to 36.4~km, yielding angular
resolutions of $\sim$42~mas and $\sim$80~mas at $Q$-band and $K$-band, respectively. Twenty-seven antennas were nominally present in the array for all of the VLA observations, although on each date, data from several antennas had to be fully or partially flagged owing to technical issues. 

In both 2023 and 2024 the $Q$-band observations were performed with the 3 bit
observing mode and dual circular polarizations. The WIDAR
correlator was configured with four baseband pairs tuned to
contiguously cover a total bandwidth of $\sim$8~GHz, centered
near 44~GHz. Each baseband pair contained 16
subbands, each with a bandwidth of 128~MHz and 128
spectral channels. The $K$-band observations employed an analogous setup, but with a center frequency of $\sim$22 GHz. Data in both bands
were recorded with 2~s time resolution.

In 2023 two epochs of observations were obtained for both $K$ and $Q$ bands, separated by 9--10 days (see below and Table~\ref{tab:newobs}). 
On 2023 September 9, observations in $Q$ and $K$ band were interleaved during a 4.0~hr session spanning approximately 13:00--17:00~UTC. Total integration times on Betelgeuse were
59.6 minutes in $Q$ band and 48.3 minutes in $K$ band.
The observations were carried out during mid-to-late morning local time and the elevation of the star ranged from $\sim$40$^{\circ}$ to 65$^{\circ}$. Weather conditions were partly cloudy and dry with wind speeds $<$1~m s$^{-1}$  and rms
atmospheric phase fluctuations of $\sim3^{\circ}$--5$^{\circ}$ during the bulk of the session.

Owing to scheduling constraints, the second 2023 epochs in each of the two bands were obtained in separate observing sessions, 
separated by one day (2023 September 19 and 20 for $K$ and $Q$ band, respectively; see Table~\ref{tab:newobs}). The $K$-band observations on September 19 were carried out from 13:35--15:11~UTC under clear skies with wind speeds  of $\le$0.7~m~s$^{-1}$. The 
atmospheric phase fluctuations ranged from 5.8$^{\circ}$--7.5$^{\circ}$ and the elevation of the star ranged from $\sim52^{\circ}$ to $\sim63^{\circ}$. The total on-source integration time was 61.4 minutes.
The $Q$-band observations on September 20 spanned approximately 11:15--13:00~UTC, corresponding to morning local time. The star was observed between elevations of $\sim$55$^{\circ}$ to 63$^{\circ}$ for a total integration time of 42.3~min. Weather was partly cloudy with wind speeds $\lsim$1~m~s$^{-1}$ and the rms phase fluctuations were 2$^{\circ}$--5$^{\circ}$. 

The 2024 VLA observations were carried out on 2024 October 23. $K$- and $Q$-band observations were interleaved during a single observing block that ran from approximately 08:00--12:00~UTC. Conditions were clear, with wind speeds $\le$1.4~m~s$^{-1}$ and rms atmospheric phase fluctuations $\lsim2^{\circ}$. The
elevation range of Betelgeuse spanned from $\sim47^{\circ}$ to 63$^{\circ}$. Total integration times on the star in each band were 59.6~min in $Q$ band and 48.4~min in $K$ band.

In 2026, observations were obtained in $K$-band only, spanning 23:26--1:13~UT on April 4--5. Conditions were partly cloudy
with wind speeds of $\sim$5--7~m~s$^{-1}$ and rms phase fluctuations of 6$^{\circ}$--8$^{\circ}$. The total integration
time on Betelgeuse was 60.7~min and the source's elevation spanned $\sim60^{\circ}$ to 65$^{\circ}$.

During all VLA observing sessions antenna pointing corrections were
evaluated hourly using observations of a strong point source in $X$ band ($\nu\sim$8~GHz). 
In all cases, fast switching between Betelgeuse and two neighboring 
gain calibrators (J0532+0732 and J0552+0313) was used for initial calibration of the atmospheric phases. The sequences and cadences used were identical to those described
in \cite{Matthews2022}. The quasar 3C48 was observed in both the $K$ and $Q$
bands to calibrate the absolute flux density scale. 

\subsection{ALMA}
\subsubsection{Bands 3 and 4}
Observations of Betelgeuse were obtained with ALMA in Band~3 ($\nu$=107~GHz; $\lambda$2.8~mm) and Band~4 ($\nu$=136~GHz; $\lambda$2.2~mm) on 2021 September 11 and 12, respectively using a hybrid long baseline configuration (C43-9/10) as part of program 2019.1.01098.S.
Baseline lengths ranged from 178~m to 16.196~km. The Band~3 observations used an array
of 44 12~m antennas.  Weather conditions were dry, with 0.62~mm of precipitable water vapor (PWV). Mean wind speeds were
$\sim$8.4~m s$^{-1}$ and mean rms phase fluctuations were 77$\mu$m. The Band~4 observations employed 45 12~m antennas in the array.
Weather conditions for the Band~4 session included PWV$\sim$1.4~mm, average wind speeds of $\sim$9.1~m s$^{-1}$, and rms phase fluctuations $\sim$175$\mu$m.

The observations in Band~3 were taken with dual linear polarizations 
using the Baseline Correlator configured with four spectral windows centered at 100.5, 102.5, 112.5, and 114.5~GHz, 
respectively, each with an effective bandwidth of
1.875~GHz, spanned by 1920 spectral channels. The Band~4 observations used a similar set-up, but with the spectral windows centered at 129.0, 130.2, 141.0, and 143.0~GHz. 
For both bands J0510+1800 was used as a bandpass and absolute flux density calibrator
and J0552+0313 (at a projected separation of
\ad{4}{2} from Betelgeuse)
was used to calibrate the time-dependent complex gains. 
A total of 45.0 minutes of integration time on Betelgeuse was obtained in each of the two bands.

\subsubsection{Bands 6, 7, and 8}
ALMA observations in Band~6 ($\nu$=223.7~GHz; $\lambda$1.3~mm), Band 7 ($\nu$=337.8~GHz; $\lambda$0.89~mm), and Band 8 ($\nu$=485.2~GHz; $\lambda$0.62~mm) 
were carried out in 2023 as part of project 2022.A.00026.S 
using the long-baseline C-10 and C-9 configurations, which provided baselines ranging from 230~m to 16.196~km. Two epochs of observations 
were obtained in both Band 6 and Band 7, while
a single epoch was observed for Band~8. The observing dates
and times for each band are summarized in Table~\ref{tab:newobs}. 
There were respectively, 45 and 49 12~m antennas in the array for the Band~6 epochs, 46 and 49 antennas for Band~7, and 39 antennas with useful data for Band~8.
For each of the bands the configuration of the Baseline Correlator included four 1.875~GHz-wide spectral windows. These were centered at 214.769, 217.073, 230.266, 231.966~GHz for Band~6, 331.052, 332,493, 343.159, 345.050~GHz for Band~7, and 478.127, 480.127, 490.231, 492.126~GHz in Band~8, respectively. 
The total on-source integration times were 30.8~min in Band~6 (per epoch), 90.8~min in Band 7 (per epoch), and 47.4~min in Band 8.
Weather conditions for Band~6 epoch~1 (epoch~2) were PWV$\sim$0.3~mm (1.1~mm), mean wind speed $\sim$4.9~m s$^{-1}$ (7.8~m s$^{-1}$), and phase rms $\sim$58.7~$\mu$m (117.0~$\mu$m). For the two Band~7 epochs, PWV was $\sim$0.3~mm (1.0~mm), 
wind speed $\sim$5.5~m s$^{-1}$ ($\sim$2.9~m $^{-1}$), and phase rms was $\sim$118.7~$\mu$m ($\sim$54.6~$\mu$m). For Band~8, the PWV was $\sim$0.4~mm, wind speed $\sim$3.2~m s$^{-1}$, and phase fluctuations $\sim65.2\mu$m.

J0510+1800 was used as a bandpass and absolute flux calibrator for all three bands. J0552+0313 (at a projected separation of \ad{4}{22} from Betelgeuse) was used as a gain calibrator for Bands 6 and 7,
while J0532+0732 (at a projected separation of \ad{5}{58}) was used as a gain calibrator for Band~8.

\section{Data Reduction, Calibration, and Imaging\protect\label{reduction}}
\subsection{VLA}
Data reduction and calibration for the VLA observations were performed using the Astronomical Image Processing
System (AIPS; \citealt{Greisen2003}), following procedures analogous to those described in \cite{Matthews2022}.
Residual instrumental delays
were corrected via fringe fitting to a 1-minute segment of data from the calibrator J0532+0732. 
J0532+0732 was also used as the bandpass calibrator, assuming
a spectral index of $\alpha\approx$0 \citep{Healey2007}. The absolute flux density scale was calibrated using 3C48 and adopting the standard coefficients of \cite{PerleyButler2017}.
Typical uncertainties in the absolute flux density scale at the VLA for our observing frequencies are 
$\sim$10--15\%\footnote{{\url{https://science.nrao.edu/facilities/vla/docs/manuals/oss/performance/fdscale}}}.

Following these calibration steps, the phase solutions for each of the Betelgeuse data sets were further improved using 1--3 iterations of phase-only self-calibration (self-cal). Solution intervals were 20~s and 120~s for the $Q$- and $K$-band data, respectively.

\subsection{ALMA}
\subsubsection{Bands 3 and 4\protect\label{B3-4}}
The raw $u$-$v$ data for ALMA project 2019.1.01098.S were retrieved from the ALMA archive and an initial series of 
standard calibration tasks (including bandpass, gain, and absolute flux calibration) was executed using the Common Astronomy Software Applications (CASA) package \citep{CASA2022}, version 6.4.1.12, along with the standard
scripts provided by the Joint ALMA Observatory (JAO). No continuum subtraction or self-calibration (self-cal) was performed at this stage. Following these initial calibration steps,
the data for Betelgeuse were split out and exported from CASA as UVFITS files and imported into AIPS for additional calibration and processing.

Owing to the proper motion of Betelgeuse, the star was slightly offset from the
field-of-view center. After shifting the star to the field center, the calibrated data were examined to identify spectral line emission and
absorption features and the corresponding spectral channels were flagged. 
(Analysis of some of these spectral line data are presented elsewhere; see \citealt{Dent2024}, \citeyear{Dent2026}).
Subsequently the data were spectrally averaged to a channel
spacing of 125.0~MHz and self-cal was performed to improve the gain calibration. For Band~3, two iterations of phase-only self-cal
were done using 10~s solution intervals, followed by a single iteration of amplitude and phase self-cal. 
For Band~4, four iterations
of phase-only self-cal were carried out with 4-s solution intervals, followed by a single iteration of amplitude and phase self-cal.
The measured correlated flux density of star increased by $\sim$2\% in Band~3 and $\sim$36\% in Band~4, respectively, following self-cal.

\subsubsection{Bands 6, 7, and 8}
The processing of the ALMA data from project 2022.A.00026.S was carried out in a similar manner to the ALMA Band 3 and 4 data (Section~\ref{B3-4}). For each band and observing epoch, 
standard calibrations (bandpass, gain, absolute flux) were applied
to the raw $u$-$v$ data using a JAO-supplied script running in CASA. No continuum subtraction or
self-cal was performed during this initial processing. 
Subsequently the Betelgeuse data were split out, converted from CASA measurement sets to UVFITS format, and imported into AIPS. 

After flagging spectral channels containing signatures of spectral lines in emission or absorption, the Betelgeuse data were averaged to a frequency resolution of 125.0~MHz.
Next, self-cal was performed for each band and each epoch of observations. For the first Band~6 epoch, 6 iterations of phase-only self-cal were carried out with 
a solution interval of 2~s, followed by a single iteration of amplitude and phase self-cal with a 5~min solution interval. These data were then imaged and used as a starting model to self-cal the second
Band~6 epoch, which had significantly poorer phase stability than epoch~1. For Band~6, epoch 2, 3 iterations of phase-only self-cal were followed by a single iteration of amplitude and phase
self-cal. 

For Band~7, epoch 1, 4 iterations of phase-only self-cal (2~s solution interval) were followed by a single iteration of amplitude and phase self-cal. These data were then imaged and used as a starting model for Band~7, epoch 2. For epoch 2, a single iteration of phase-only self-cal was performed, followed by a single iteration of amplitude and phase self-cal.
For Band~8, two iterations of phase-only self-cal were performed
using 10~s solution intervals, followed by a single iteration of amplitude and phase self-cal.

As a quality assurance check, Post-pipeline processing of the Betelgeuse data in all three bands was also performed independently within CASA (see \citealt{Dent2024}, \citeyear{Dent2026}). The results were generally 
of comparable quality, with no statistically significant differences in measured parameters for the star. We use the AIPS-processed data for all further imaging and analysis presented here.

%
\begin{table}
\caption{Summary of New Betelgeuse Observations}
\centering
\label{tab:newobs}
\begin{tabular}{lcccccc}
\hline
Date (UTC) & Band & $\nu_{0}$(GHz) & Wavelength (mm) & Telescope & Baseline Lengths (km) & $\theta_{\rm LRS}$ (arcsec)\\
\hline
2023-09-09 & $K$ band & 22.0 & 13.6 & VLA & 0.68--36.4 & \as{2}{4}\\
2023-09-19 & $K$ band & 22.0 & 13.6 & VLA & 0.68--36.4 & \as{2}{4}\\
2024-10-23 & $K$ band & 22.0 & 13.6 & VLA & 0.68--36.4 & \as{2}{4}\\
2026-04-04 & $K$ band & 22.0 & 13.6 & VLA & 0.68--36.4 & \as{2}{4} \\
2023-09-09 & $Q$ band &  44.0 & 6.8 & VLA & 0.68--36.4 & \as{1}{2}\\
2023-09-20 & $Q$ band & 44.0 & 6.8 & VLA & 0.68--36.4 & \as{1}{2}\\
2024-10-23 & $Q$ band & 44.0 & 6.8 & VLA & 0.68--36.4 & \as{1}{2}\\
2021-09-11 & Band 3 & 107.5 & 2.8 & ALMA & 0.178--16.196 & \as{0}{50}\\
2021-09-12 & Band 4 & 135.6 & 2.2 & ALMA & 0.178--16.196 & \as{0}{33}\\
2023-08-03 & Band 6 & 223.7 & 1.3 & ALMA & 0.230--16.196 & \as{0}{22}\\
2023-08-27 & Band 6 & 223.7 & 1.3 & ALMA & 0.230--16.196 & \as{0}{22}\\
2023-08-03 & Band 7 & 337.8 & 0.89 & ALMA & 0.230--16.196 & \as{0}{14} \\
2023-08-04 & Band 7 & 337.8 & 0.89 & ALMA & 0.230--16.196 & \as{0}{14}\\
2023-08-04 & Band 8 & 485.2 & 0.62 & ALMA & 0.230--16.196 & \as{0}{11}\\
\hline
\end{tabular}
\flushleft Quoted frequencies ($\nu_{0}$) correspond to the mean observed frequency in each receiver band.
$\theta_{\rm LSR}$ is an estimate of the largest recoverable angular scale of the interferometer.
\end{table}
%
\begin{table}
\caption{VLA Calibration Sources}
\centering
\label{tab:vlacalibration}
\begin{tabular}{lrrccl}
\hline
Source & $\alpha$(J2000.0) & $\delta$(J2000.0) & Flux Density (Jy) & $\nu$ (GHz) & Date(s)\\
\hline
3C48$^{a}$ & 01 37 41.2994 & 33 09 35.133 & 0.6134$^{*}$ & 44.0  & 2023-09-09, 2023-09-20, 2024-10-23\\
                          ... & ... & ... & 1.2194$^{*}$ & 22.0 & 2023-09-09, 2023-09-19, 2024-10-23, 2026-04-04\\

J0532+0732$^{b}$ & 05 32 38.9985 & 07 32 43.346 & 1.926$\pm$0.039 & 44.0 & 2023-09-09\\
    ...          & ...           & ...          & 1.800$\pm$0.039 & 44.0 & 2023-09-20\\
          ...        &     ...       &  ...     & 1.4004$\pm$0.038 &44.0 & 2024-10-23\\
      ...        &     ...       &  ...         & 2.233$\pm$0.012 & 22.0 & 2023-09-09  \\
           ...          & ...   & ...           & 2.187$\pm$0.014  & 22.0 & 2023-09-19\\
           ...     &    ... &    ...            & 1.411$\pm$0.012 & 22.0 & 2024-10-23\\ 
           ...      &   ... &    ...            & 1.231$\pm$0.006 & 22.0 & 2026-04-04 \\
J0552+0313$^{c}$ & 05 52 50.1015 & 03 13 27.243 & 0.444$\pm$0.009 & 44.0 & 2023-09-09\\
    ...          & ...           & ...          & 0.498$\pm$0.010 & 44.0 & 2023-09-20\\
    ...          & ...           & ...          & 0.507$\pm$0.014 & 44.0 & 2024-10-23\\
    ...         & ...           & ...           & 0.510$\pm$0.003 & 22.0 & 2023-09-09\\
      ...         & ...           & ...         & 0.537$\pm$0.004 & 22.0 & 2023-09-19\\
      ...         & ...           & ...         & 0.597$\pm$0.005 & 22.0 & 2024-10-23\\
      ...         & ...           & ...         & 0.671$\pm$0.003 & 22.0 & 2026-04-04\\
    \hline
\end{tabular}
\flushleft
Units of right ascension are hours, minutes, and
seconds, and units of declination are degrees, arcminutes, and
arcseconds. Explanation of columns: (1) source name; (2) \& (3) right
ascension and declination (J2000.0); (4) flux density in Jy at the frequency indicated in column 5;
(5) frequency at which the flux density in the fourth column was computed.

$^{*}$Adopted value, calculated at the frequency in column~5 and the  coefficients from \cite{PerleyButler2017}.
  
$^{a}$Flux density calibrator\\
$^{b}$Complex gain and bandpass calibrator\\
$^{c}$Complex gain calibrator\\

\end{table}
%
\begin{table}
\caption{ALMA Calibration Sources}
\centering
\label{tab:almacalibration}
\begin{tabular}{lllccc}
\hline
Source & $\alpha$(J2000.0) & $\delta$(J2000.0) & Flux Density (Jy) & $\nu$ (GHz) & Date(s)\\
\hline
J0510+1800$^{a}$ & 05:10:02.36913 & +18:00:41.5817 & 1.84$^{*}$ & 100.5 & 2021-09-11\\
                          ... & ... & ...          & 1.59$^{*}$ & 129.0 & 2021-09-12\\

J0552+0313$^{b}$ &  05:52:50.10146& +03:13:27.2445 & 0.356$\pm$0.018 & 100.5 & 2021-09-11\\
    ...          & ...           & ...             &  0.309$\pm$0.021 & 129.0 & 2021-09-12\\
    ...          & ...           & ...             & 0.161$\pm$0.008 & 214.8 & 2023-08-03\\
    ...         & ...           & ...             &  0.168$\pm$0.008     & 214.8 & 2023-08-27\\
    ...          & ...           & ...             & 0.104$\pm$0.010  & 331.1 & 2023-08-03 \\
    ...          & ...           & ...             & 0.109$\pm$0.013       & 331.1 & 2023-08-04\\

J0532+0732$^{b}$  & 05 32 38.988 & 07 32 43.345     & 0.393$\pm$0.084 & 478.2 &2023-08-01 \\    
    \hline
\end{tabular}
\flushleft
Column definitions are as in Table~\ref{tab:vlacalibration}.

$^{*}$Adopted value, calculated at the frequency in
  column~5 and assuming a spectral index of $\alpha$=$-0.534$. 

$^{a}$Flux density and bandpass calibrator\\
$^{b}$Complex gain calibrator\\

\end{table}

\subsection{Imaging\protect\label{imaging}}
\begin{figure*}
\begin{center}
  \includegraphics[width=0.4\textwidth,angle=0]{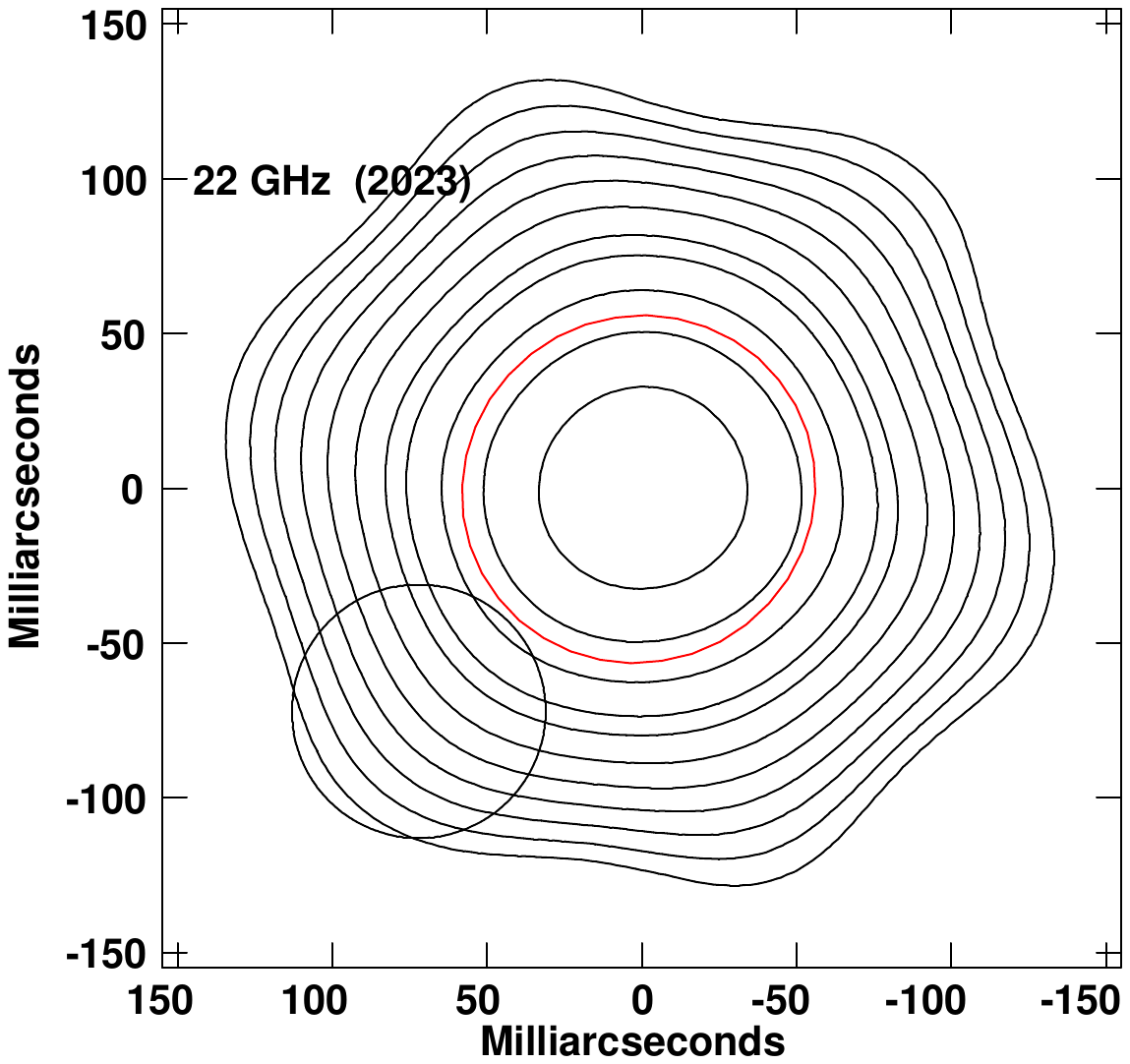}\hspace{-2.6cm}
	\includegraphics[width=0.4\textwidth,angle=0]{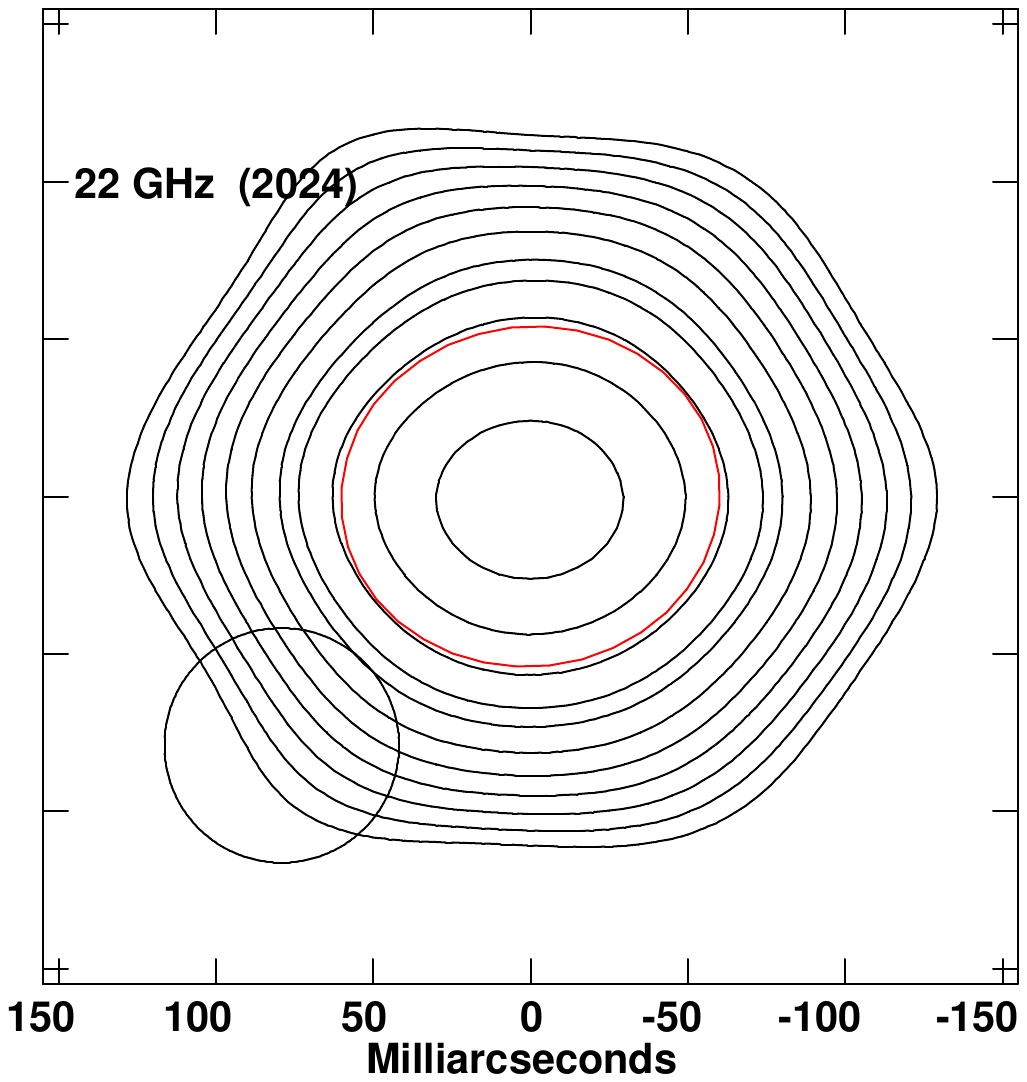}\hspace{-2.6cm}\includegraphics[width=0.4\textwidth,angle=0]{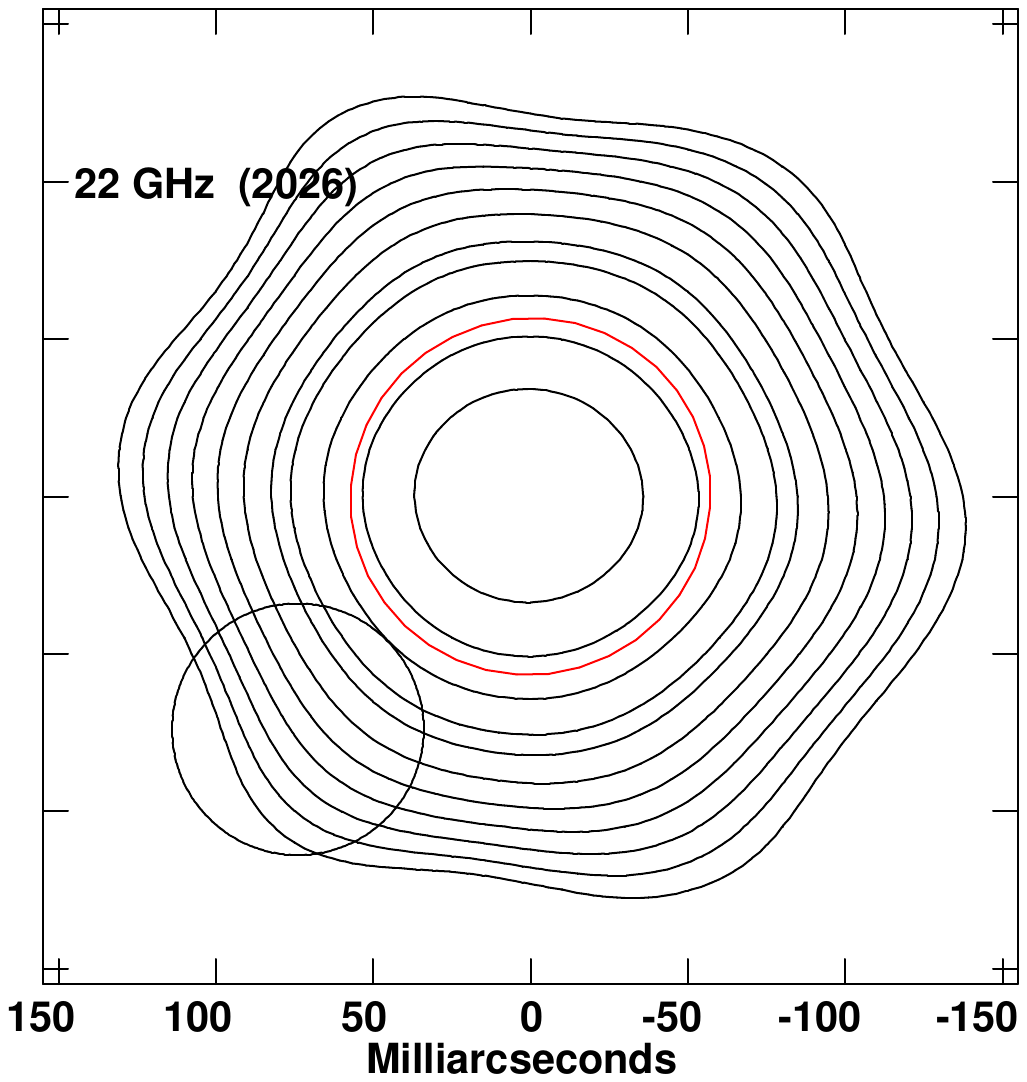}
    \includegraphics[width=0.4\textwidth,angle=0]{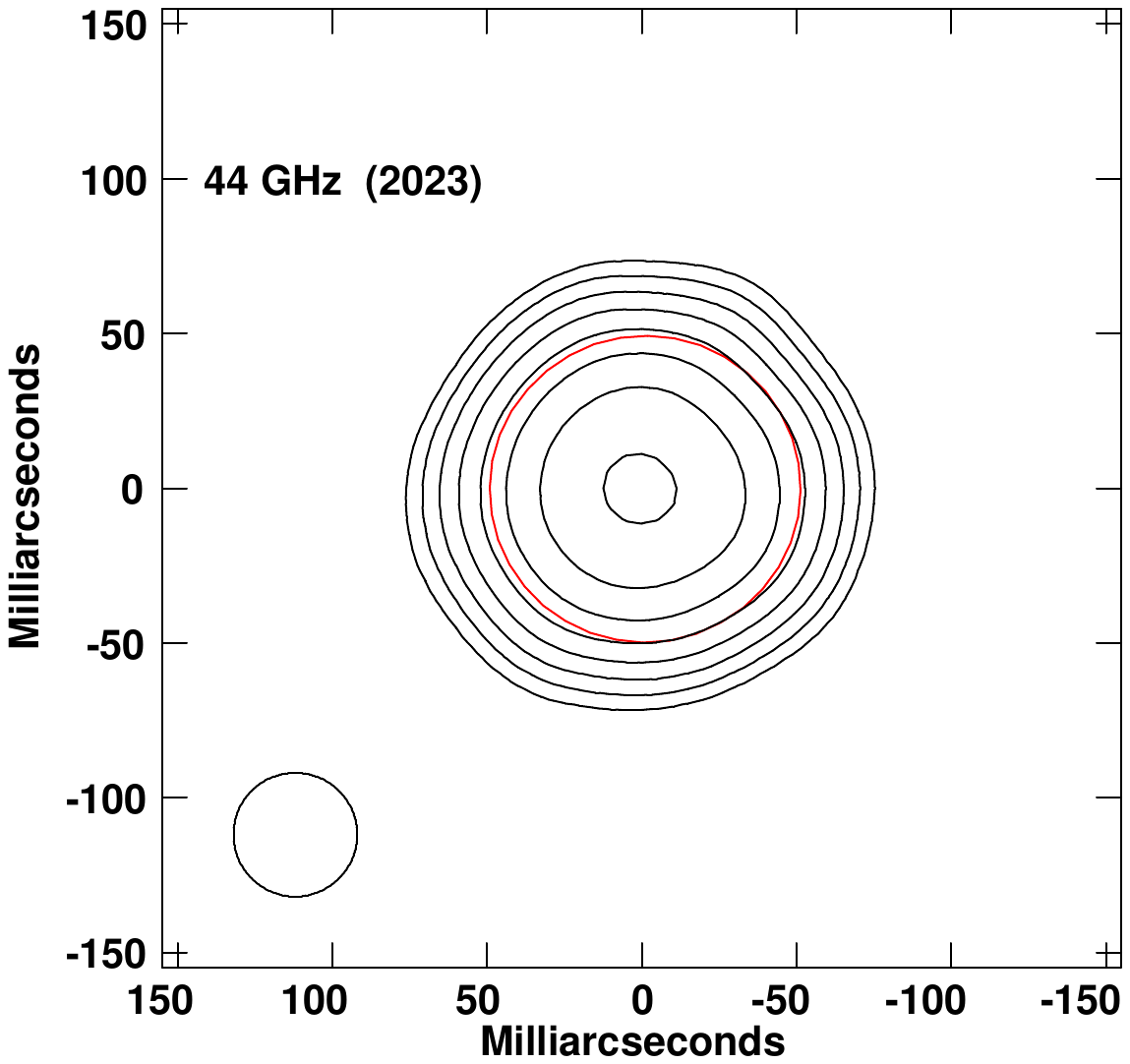}\hspace{-2.6cm}
   \includegraphics[width=0.4\textwidth,angle=0]{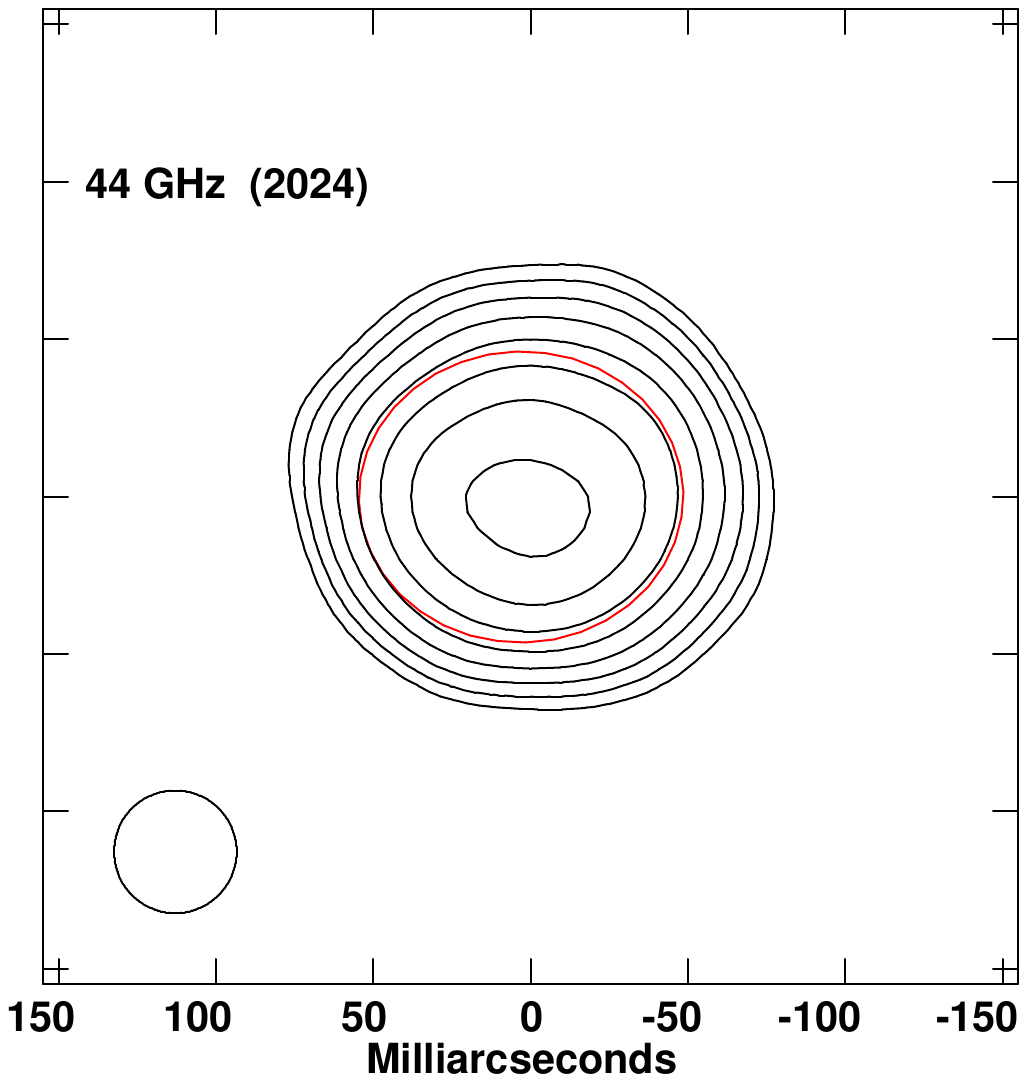} 
    \caption{VLA contour images of Betelgeuse from 2023, 2024, and 2026. 22~GHz data are shown in the top row and 44~GHz data in the bottom row. North is on top and east is to the left. The 2023 images use combined data from the two observing epochs (see Table~1). All images were made using robust ${\cal R}$=0 weighting and a circular restoring beam with dimensions equal to the geometric mean of the dirty beam dimensions given in Table~\ref{tab:newimages}. The restoring beam is indicated in the lower left corner of each panel. 
    Contour levels are spaced by $\sqrt{2}$: 8.8$\times$($-20$[absent], 20, 28.3,...565.7) $\mu$Jy beam$^{-1}$ for the 22~GHz data and 34.0$\times$($-20$[absent], 20, 28.3,...226.3) $\mu$Jy beam$^{-1}$ for the 44~GHz data.  For ease of comparison, the images in each respective band are contoured identically. The lowest contours correspond to $\pm$20$\sigma$, where $\sigma$ is the rms noise of the 2024 data, as given in Table~\ref{tab:newimages}. The overplotted red ellipses indicate the dimensions of the best-fitting uniform elliptical disk model from Table~\ref{tab:measurements} (see Section~\ref{unidisk}).  The ``hexagon'' shape of the 22~GHz images is an artifact of the VLA point spread function that manifests as result of the stellar disk being only slightly larger than the FWHM of the synthesized beam.  
    }
    \label{fig:vlakntrmaps}
    \end{center}
\end{figure*}

\begin{figure*}
\begin{center}
    \includegraphics[width=0.4\textwidth,angle=0]{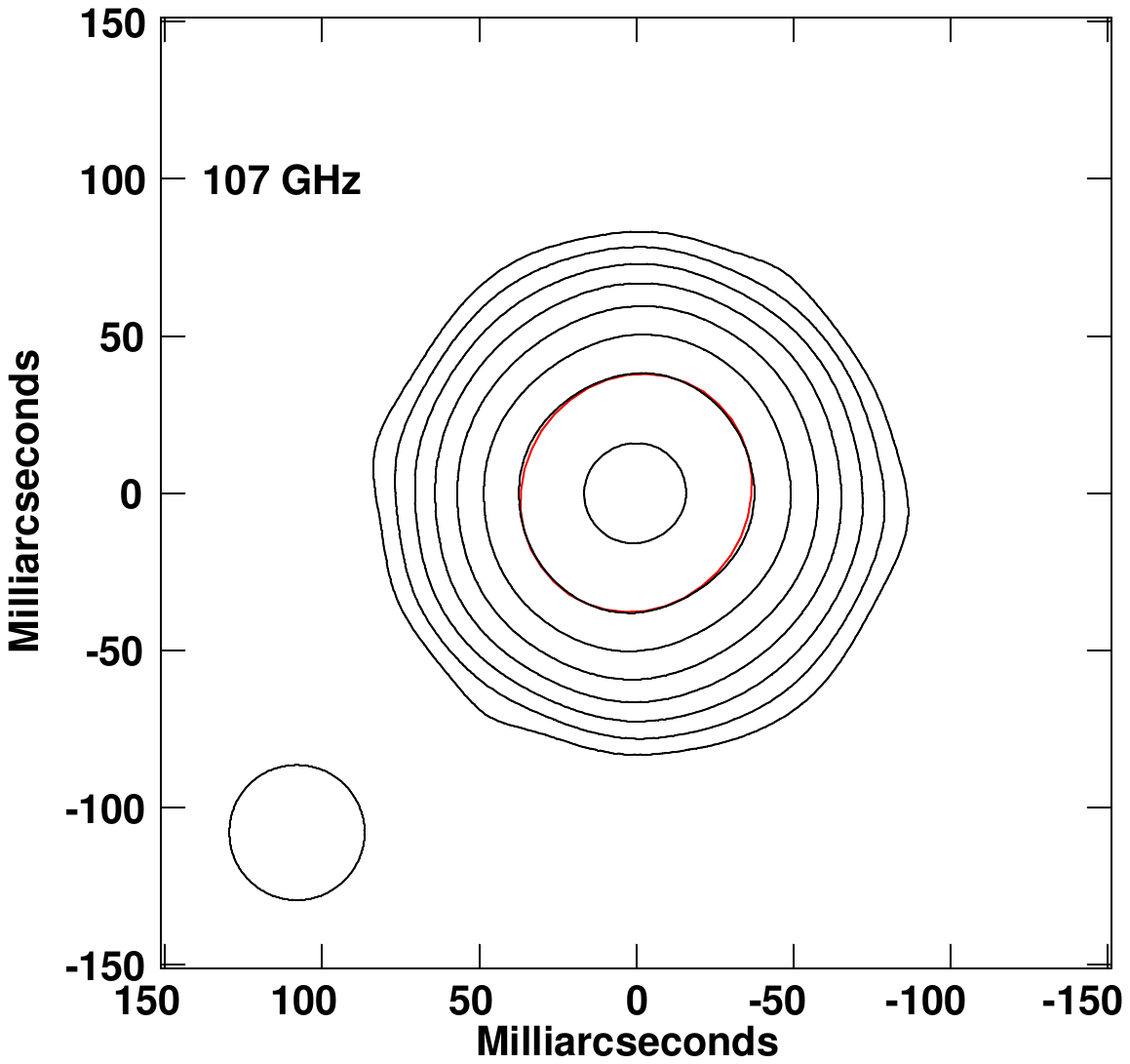}\hspace{-2.6cm}
    \includegraphics[width=0.4\textwidth,angle=0]{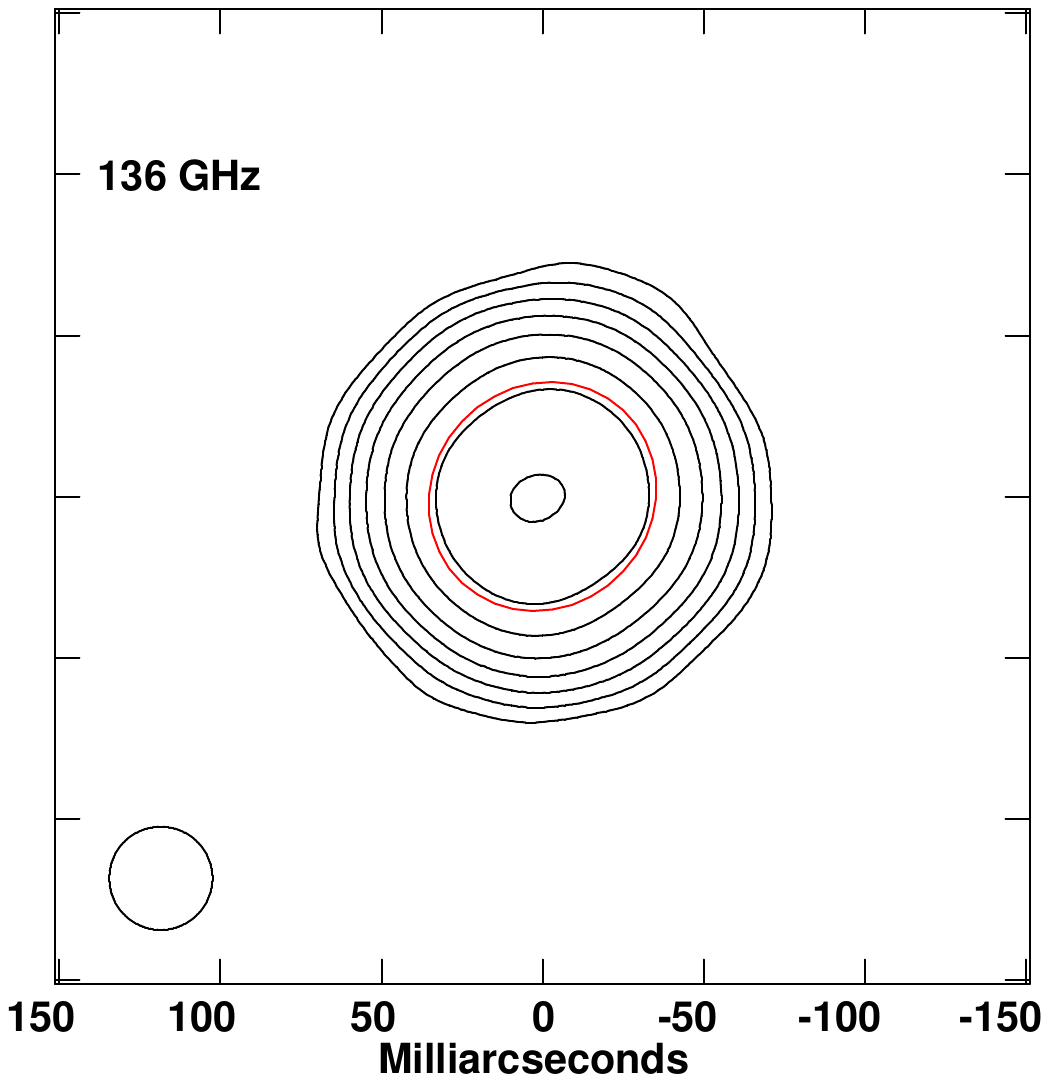}
    \caption{Contour images of Betelgeuse from 2021 ALMA observations at 107~GHz (left) left and 136~GHz (right). The images were made using robust ${\cal R}$=0 weighting and a circular restoring beam with dimensions equal to the geometric mean of the dirty beam dimensions (see Table~\ref{tab:newimages}). Contour levels are spaced by factors of 2 and are 15.0$\times$($-20$[absent], 20, 40,...2560) $\mu$Jy beam$^{-1}$ for both bands. The lowest contours are $\sim\pm$20$\sigma$, where $\sigma$ is the rms noise given in Table~\ref{tab:newimages}. The restoring beam is indicated in the lower left corner of each panel. The red ellipses indicate the dimensions of the best-fitting uniform elliptical disk model from Table~\ref{tab:measurements} (see Section~\ref{unidisk}). 
    }
    \label{fig:almakntrmaps1}
    \end{center}
\end{figure*}

\begin{figure*}
\begin{center}
    \includegraphics[width=0.4\textwidth,angle=0]{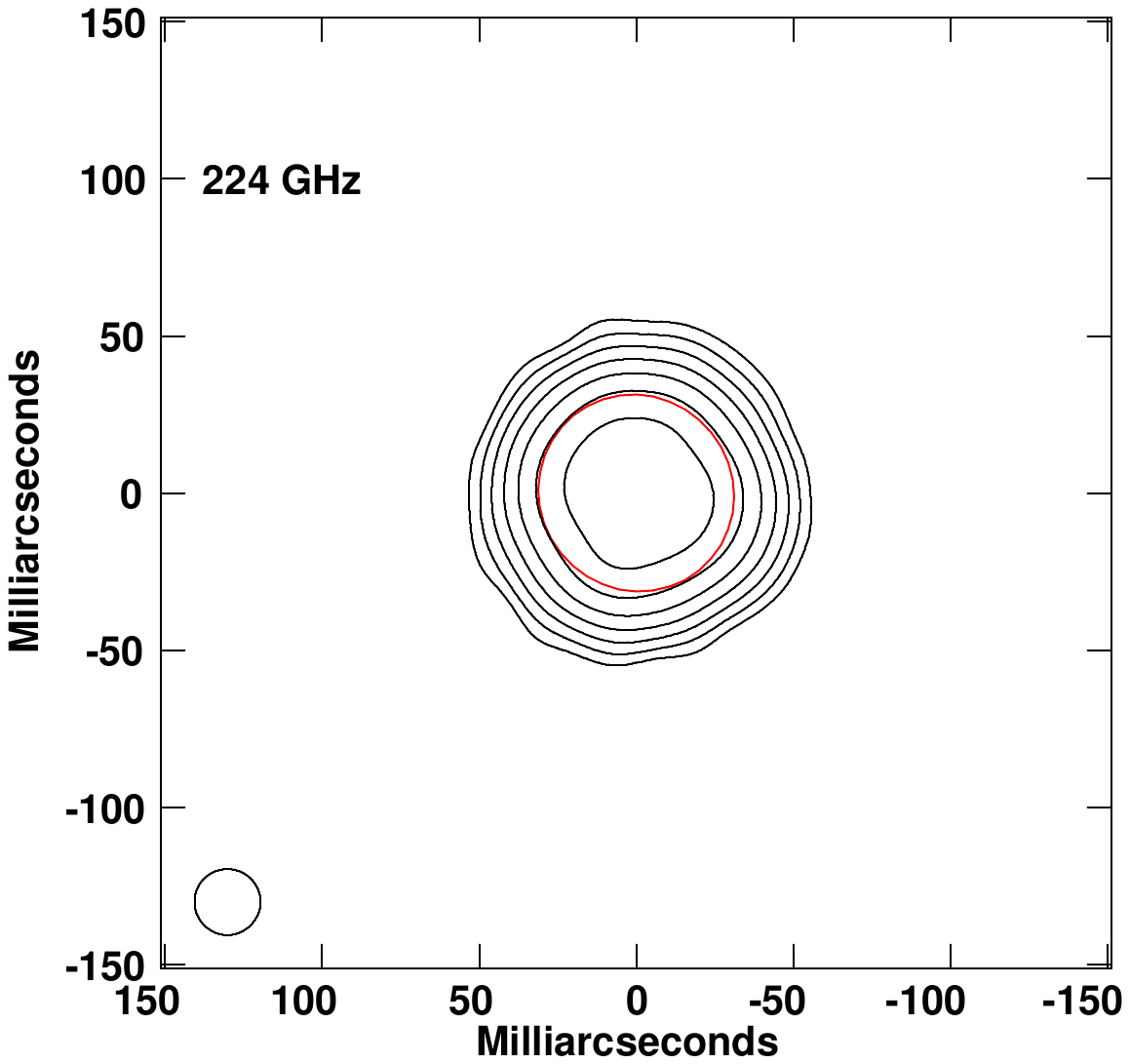}\hspace{-2.6cm}
    \includegraphics[width=0.4\textwidth,angle=0]{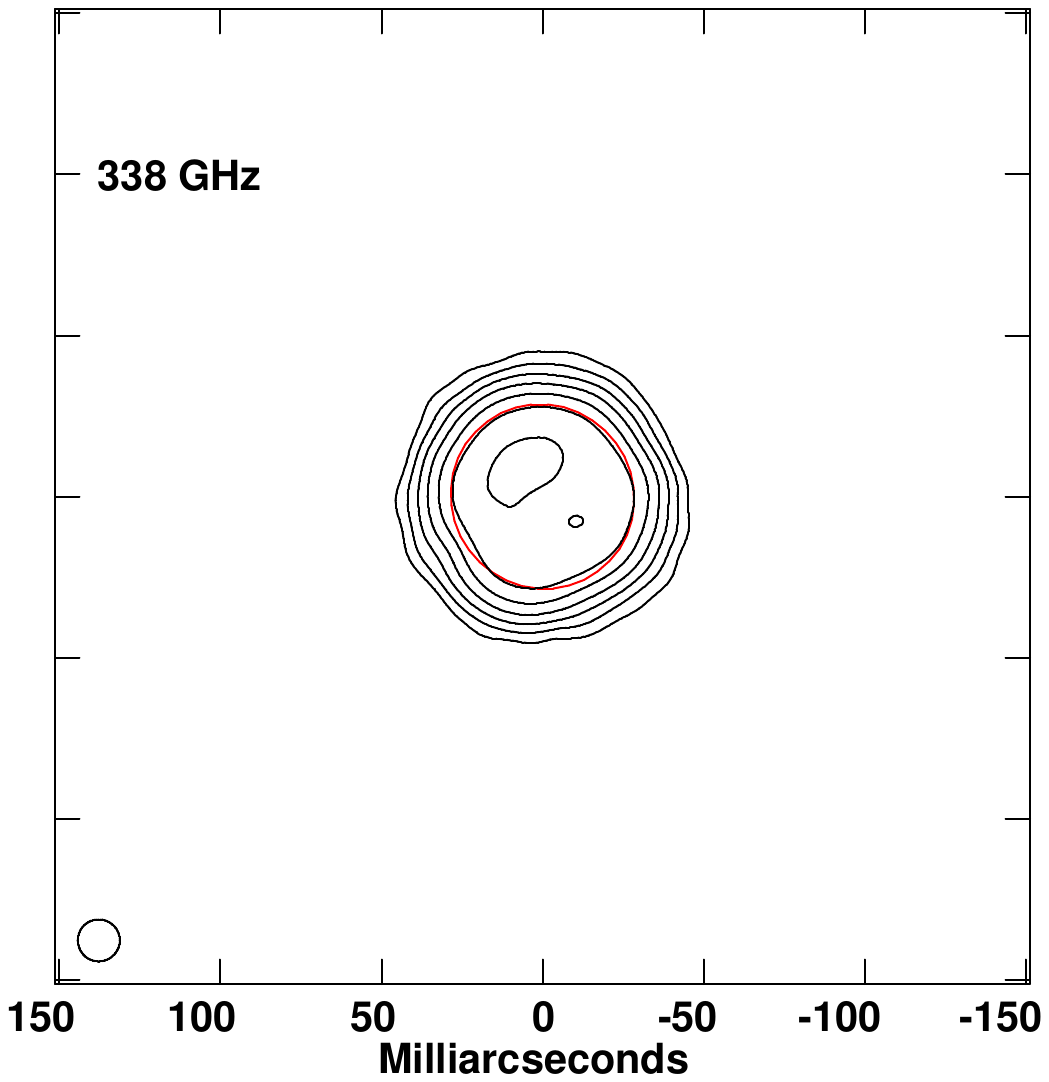}\hspace{-2.6cm}
    \includegraphics[width=0.4\textwidth,angle=0]{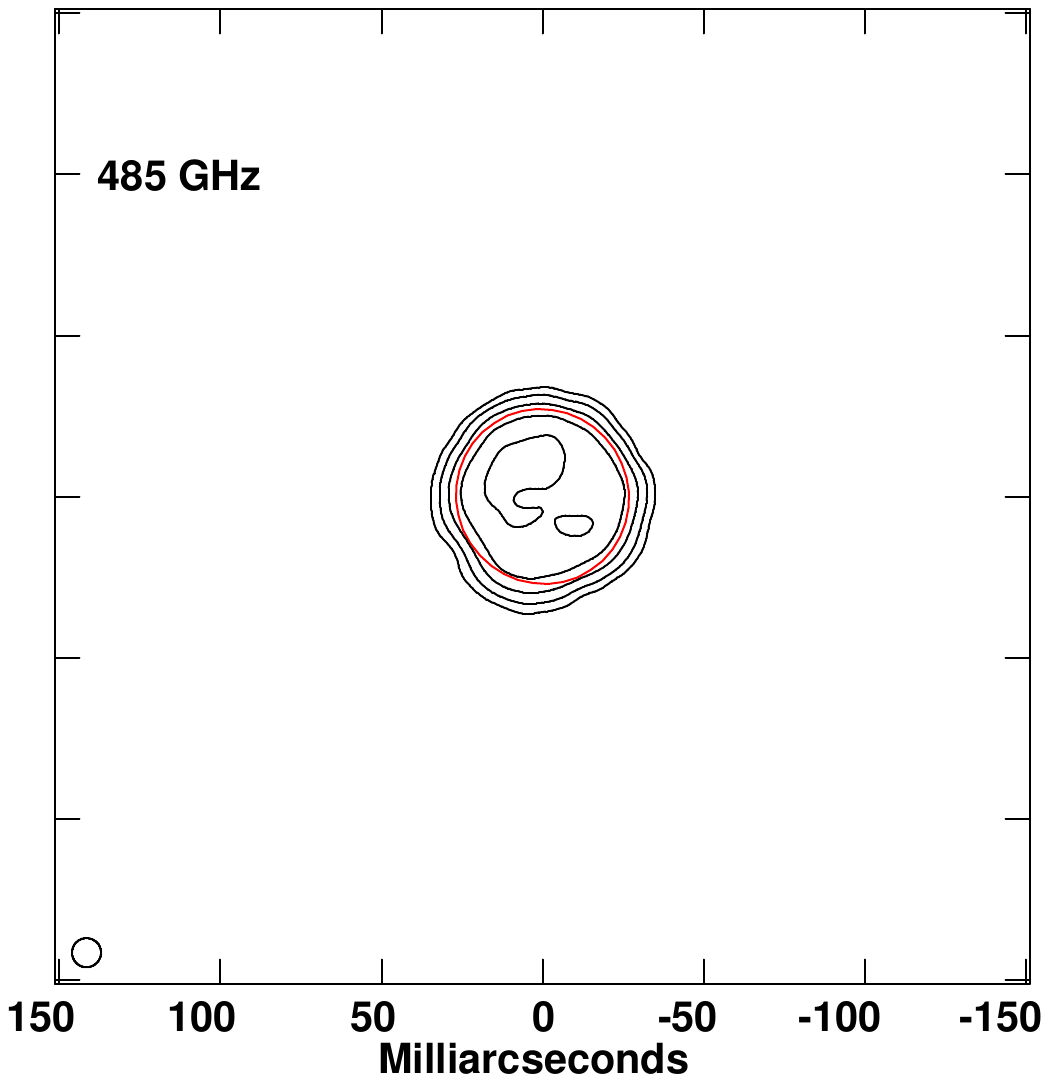}
    \caption{As in Figure~\ref{fig:almakntrmaps1}, but for  2023 ALMA data: 224~GHz (left); 338~GHz (center); 485~GHz (right).  
   The contour levels are spaced by factors of 2: 26.2$\times$($-20$[absent], 20, 40,...1280) $\mu$Jy beam$^{-1}$ at 224~GHz; 29.3$\times$($-20$[absent], 20, 40,...2560) $\mu$Jy beam$^{-1}$ at 338~GHz; and 135.0$\times$($-20$[absent], 20, 40,...260) $\mu$Jy beam$^{-1}$ at 485~GHz.}
    \label{fig:almakntrmaps2}
    \end{center}
\end{figure*}

Using the fully calibrated VLA and ALMA data we produced images of Betelgeuse in each of our observing bands
using {\sc CLEAN} deconvolution as implemented
in the AIPS {\sc IMAGR} task (Figures~\ref{fig:vlakntrmaps}, \ref{fig:almakntrmaps1}, and \ref{fig:almakntrmaps2}). 
Multi-scale CLEAN was used for the ALMA data, while standard CLEAN was used for the VLA data. The images were all produced using 
robust weighting with ${\cal R}$=0 and a circular restoring beam 
with a FWHM equal to the geometric mean of the dimensions of the dirty beam (see Table~\ref{tab:newimages}). 

The images demonstrate that Betelgeuse is spatially resolved in all of our observing bands and they clearly illustrate the systematic decrease in size of
the radio disk with increasing frequency. This is further quantified in Section~\ref{analysis}.

Because the disk of the star is only marginally resolved at our two lowest observing frequencies (22 and 44~GHz), the corresponding images cannot place strong constraints on the possible presence of surface features or brightness irregularities. Nonetheless, it is worth noting that the consistently smooth and featureless appearance of the stellar disk in these bands across multiple epochs (Figure~\ref{fig:vlakntrmaps}; see also \citealt{Matthews2022}) contrasts with the earlier study of \cite{Lim1998}, who reported evidence for a giant convective cell and  an irregular atmospheric shape for Betelgeuse based on a 44~GHz VLA image with comparable spatial resolution. Our latest analysis suggests that these earlier results should be interpreted cautiously, as phase calibration errors can also mimic such features; indeed we see similar morphological features in our most recent data sets prior to the application of phase self-calibration.\footnote{Phase calibration improvements from self-calibration effectively sharpen or ``focus'' radio interferometric images, enhancing image fidelity. However, this step that was not possible for legacy VLA observations of Betelgeuse owing to signal-to-noise ratio limitations.}

The ALMA 107 and 136~GHz images (Figure~\ref{fig:almakntrmaps1}) 
also appear relative featureless, but 
at the highest ALMA frequencies (224, 338, and 485~GHz; Figure~\ref{fig:almakntrmaps2}), 
additional features and brightness variations become clearly visible across the disk. In these instances, the features are robustly detected in self-calibrated images, as well as in the visibility data.
Images derived from the high-frequency ALMA data are discussed in more detail in \cite{Dent2026}. In the present paper we focus
instead of the global, multi-frequency radio properties of Betelgeuse.

%
\begin{table}
\caption{CLEAN Image Properties}
\centering
\label{tab:newimages}
\begin{tabular}{lccccccr}
\hline
Band & Date & Frequency & Wavelength & rms Noise  & Dirty Beam  & Beam PA \\
 & (UTC) & (GHz) & (mm) & ($\mu$Jy beam$^{-1}$) & (mas) & (deg)\\
\hline
\multicolumn{7}{c}{VLA}\\
\hline
$K$ & 2023-09-09 & 22.0 & 13.6 & 9.0 & 90.3$\times$74.2 & 29.6 \\
$K$ & 2023-09-19 & 22.0 & 13.6 & 7.6 & 84.5$\times$81.5 & $-2.7$\\
$K$ &  2023 combined & 22.0 & 13.6 & 6.6 & 86.2$\times$78.8 & 20.9\\
$K$ & 2024-10-23 & 22.0 & 13.6 & 8.9 & 81.3$\times$68.5 & $-12.7$\\
$K$ & 2026-04-04 & 22.0 & 13.6 & 6.4 & 87.4$\times$72.3 & $-2.7$\\

$Q$ & 2023-09-09 & 44.0 & 6.8 & 23.4 & 43.0$\times$38.7 & 17.4\\
$Q$ & 2023-09-20 & 44.0 & 6.8 & 27.8 & 47.2$\times$39.2& $-33.7$\\
$Q$ & 2023 combined & 44.0 & 6.8 & 19.3 & 42.6$\times$39.4 & $-6.5$\\
$Q$ & 2024-10-23 & 44.0 & 6.8& 34.0 &  42.3$\times$36.0 & $-13.3$\\
\hline

\multicolumn{7}{c}{ALMA}\\
\hline
3 & 2021-09-11 & 107.5 & 2.8 & 15.0& 51.1$\times$36.6 & $-33.2$ \\ 

4 & 2021-09-12 & 135.6 & 2.2 & 16.2 & 35.3$\times$29.3 & $-23.3$ \\  

6 & 2023-08-03 & 223.7 & 1.3 & 23.6 &25.0$\times$17.6 & 53.7\\ 
6 & 2023-08-27 & 223.7 & 1.3 & 31.7 & 26.4$\times$18.1 & 46.4\\ 
6 & 2023 combined & 223.7 & 1.3 & 26.2 & 25.3$\times$17.8 & 51.1 \\ 

7 & 2023-08-03 & 337.8 & 0.89 & 32.0 & 13.7$\times$12.2 & $-2.2$\\ 
7& 2023-08-04 & 337.8 & 0.89 & 44.2 & 14.2$\times$11.8 & 45.5\\
7& 2023 combined & 337.8 & 0.89 & 29.3 & 13.4$\times$12.2 & 20.0\\

8 & 2023-08-04 & 485.2 & 0.62 & 134.7 & 9.40$\times$8.64 & 1.8\\

\hline
\end{tabular}
\flushleft Quoted frequencies are at the frequency-averaged band center. The rms noise and dirty beam parameters are derived from AIPS {\sc CLEAN} images (VLA data)
or AIPS multi-scale {\sc CLEAN} images (ALMA data), produced with ${\cal R}$=0 weighting and no $u$-$v$ restrictions. PA is measured from north to east.
\end{table}

\section{Analysis}\protect\label{analysis}
\subsection{Characterization of Stellar Parameters from Visibility Data\protect\label{uvdiskfits}}
\subsubsection{Uniform Elliptical Disk Fits\protect\label{unidisk}}
Recent high-resolution imaging studies of Betelgeuse have clearly demonstrated that the radio
surface of Betelgeuse is irregular and deviates from spherical symmetry (\citealt{Matthews2022}; \citealt{OG2017}; \citealt{Dent2026}). Despite this, global characterization of the emission using simple model fits still provides a useful means to characterize mean properties of the star and to track global changes over time. 
We have therefore measured the size, shape, and flux density of Betelgeuse in each of our
observing bands using two-dimensional uniform elliptical disk fits to the visibility data with the AIPS {\sc OMFIT} task.
We
initially fit the data from each band and observing epoch separately, but in cases where multiple observations of a given band were obtained in the same year we also fit the combined data. The results are presented in Table~\ref{tab:measurements}. The tabulated error bars include
contributions from the formal fitting uncertainties as prescribed by \cite{Condon1997}, as well as band-dependent uncertainties in the absolute flux density scale.  For the latter we assume an uncertainty in the absolute flux calibration of 10\% in both of our VLA observing
bands\footnote{\url{https://science.nrao.edu/facilities/vla/docs/manuals/obsguide/topical-guides/hifreq}}, while 
for ALMA, we adopt 5\% for Bands~3, 4, 6; 7\% for Band~7; and 10\% for Band~8 \citep{Cortes2024}.

As seen from the resulting fit parameters (Table~\ref{tab:measurements}), the flux density of the star increases at higher frequencies, while the angular dimensions of the star systematically decrease. This is consistent with previous
studies (\citealt{Lim1998}; \citealt{OG2015}, \citeyear{OG2017}; \citealt{Matthews2022}) and is a hallmark of optically thick thermal emission. The inferred spectral index is discussed in Section~\ref{specindex}.

The overall shape of Betelgeuse was found to have an ellipticity $e\lsim$10\% in all observing bands. 
Nonetheless, it is noteworthy that the measured ellipticities are systematically non-zero and tend to be consistent among observations close in time.   In Section~\ref{companion} we discuss evidence that systematic changes in ellipticity occur over time and are linked with the orbit of a recently discovered binary companion.

\subsubsection{Brightness Temperature\protect\label{TB}}
Based on the best-fitting uniform elliptical disk parameters summarized in Table~\ref{tab:measurements} we have derived a disk-averaged brightness temperature, $T_{\rm B}$, from each measurement according to the relation:

\begin{equation}
T_{B}=2S_{\nu}c^2(4.25\times10^{-28})/(k\nu^{2}\pi\theta_{\rm
    maj}\theta_{\rm min})
\end{equation}

\noindent  where $S_{\nu}$ is the flux density 
  in mJy; $\nu$ is the
  observing frequency in GHz; $c$ is the speed of light in cm s$^{-1}$; $k$ is
  the Boltzmann constant in cgs units; and $\theta_{\rm maj}$ and $\theta_{\rm min}$ are the major and minor axis diameters, respectively, of the
uniform elliptical disk model fit (in  mas).
Assuming that the radio emission from Betelgeuse is optically thick, the derived brightness temperatures provide a measure 
of the mean gas (electron) temperature. 

\begin{figure*}
\begin{center}
	\includegraphics[width=15cm,angle=-180]{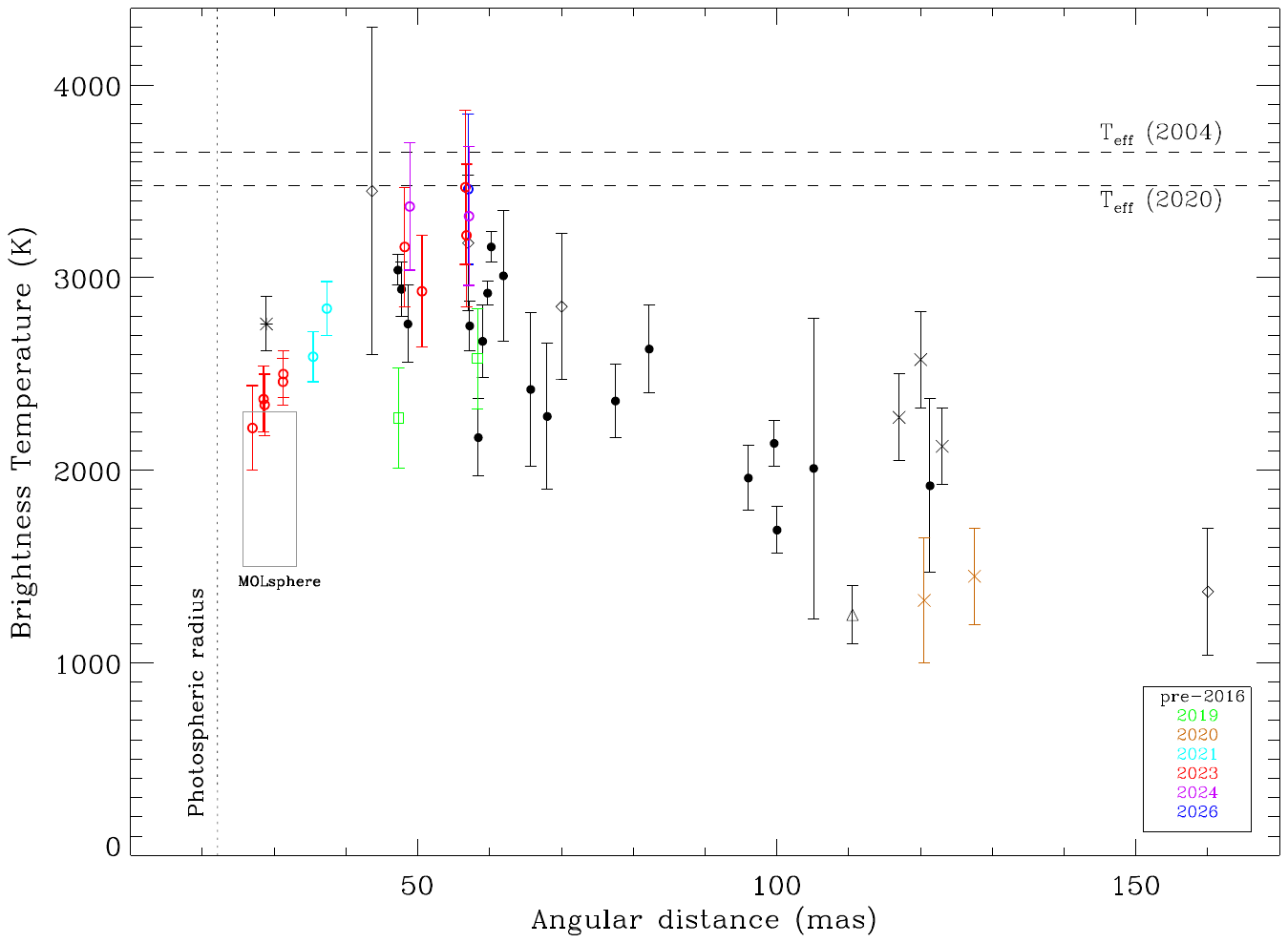}
        \caption{Mean (disk-averaged) brightness temperature (in K) versus mean angular radius (in mas), as derived from uniform elliptical disk fits to radio measurements
          of Betelgeuse. Open circles represent results from the current work (see Table~\ref{tab:measurements}), color-coded by year of observation. Green squares show the VLA measurements from \cite{Matthews2022}.  Brown crosses are from unpublished eMERLIN measurements at 6.3~GHz obtained in 2020 (A. M. S. Richards, private communication). Black points are from measurements obtained prior to 2016:  open diamonds (\citealt{Lim1998}); open triangle (\citealt{Richards2013}); filled circles: (\citealt{OG2015});
          asterisk (\citealt{OG2017}); crosses: unpublished eMERLIN measurements at 5.7~GHz obtained in 2012 and 2015 (A. M. S. Richards, private communication).
          The grey rectangle depicts the approximate range of radii and temperatures derived from previous measurements of the MOLsphere (see text for references).
          The horizontal dashed lines indicate representative photospheric effective temperatures of the star  as measured in 2004 \citep{Levesque2005} and
          in 2020 \citep{Alexeeva2021}.  The vertical dotted line indicates the angular radius of the classical photosphere as determined from the 2.2~$\mu$m measurement
          of \cite{Dyck92}.
   }
    \label{fig:tempsize}
    \end{center}
\end{figure*}

In Figure~\ref{fig:tempsize} we plot $T_{\rm B}$ as a function of the mean projected angular radius as derived from the uniform elliptical disk fits summarized in Table~\ref{tab:measurements}. Additional measurements from other authors are also overplotted, with those obtained prior to 2016 indicated with black symbols.

Although there is significant scatter in the $T_{\rm B}$ measurements from different epochs of radio observations at any given frequency, there are some noteworthy trends. First, we see that our most recent $T_{\rm B}$ measurements at both 22 and 44~GHz (which probe projected radii $45\lsim r\lsim60$~mas and are plotted on Figure~\ref{fig:tempsize} as open circles, color-coded by year) are significantly higher than the values measured in 2019 August, just prior to the onset of the Great Dimming (shown as green squares). While the 2019 $T_{B}$ measurement at 22~GHz marked a historic low \citep{Matthews2022}, the new measurements are now consistent (to within scatter) with other prior measurements at these frequencies taken over the past $\sim$30~yr. At 22~GHz, the brightness temperatures seen in our latest measurements ($T_{\rm B}\approx$3500~K)  are now comparable to the photospheric effective temperature of Betelgeuse ($T_{\rm eff}\approx3476-3650$~K; e.g., \citealt{Dyck92}; \citealt{Levesque2005}; \citealt{LMass2020}; \citealt{Alexeeva2021}). This contrasts with ultraviolet observations that probe comparable atmospheric heights, but indicate the presence of plasma at temperatures significantly higher than those in the photosphere
($\sim$4000-8000~K; e.g., \citealt{Dupree2020} and references therein). 
As discussed by \cite{Lim1998}, the dominant source of radio opacity 
changes from interactions between
electrons with neutral and molecular hydrogen for $T\lsim$4000~K to electron scattering by protons at warmer temperatures (see \citealt{RM1997}). However, since the latter produces $\sim10^{3}$ times lower opacity per particle, this implies that to account for the observed radio emission, the volume of cooler gas in the extended atmosphere must be significantly larger than that of the warmer ultraviolet-emitting material. 

One key difference compared with pre-2019 measurements is that for a given epoch of observation, $T_{\rm B}$ measured at 22~GHz is consistently comparable to or higher than the value measured at 44~GHz. Another difference is that the peak value of $T_{\rm B}$ occurs at $\gsim$60~mas, or $r\gsim2.7R_{\star}$, in contrast with earlier studies   where it was seen to occur near $r\sim$50~mas (\citealt{Lim1998}; \citealt{OG2015}).

\cite{Matthews2022} interpreted the anomalous 
2019 measurements at 22~GHz and 44~GHz as evidence of the passage through the atmosphere of a convectively induced shockwave that is now believed to be the precursor of the Great Dimming (\citealt{Kravchenko2021}; \citealt{Dupree2022}).  Since that event, optical observations have revealed
persistent changes in Betelgeuse in terms of the length of the pulsation cycle and the amplitude of radial velocity variations (Section~\ref{sec:intro}), while new ultraviolet measurements also indicate a corresponding decline in the chromospheric
density \citep{Dupree2022}. It is therefore of interest to explore whether any systematic changes are discernible at radio wavelengths compared with observations made prior to 2019.

For the 22~GHz and 44~GHz observations subsequent to the Great Dimming we find that although the {\em individual} values derived for various quantities 
(flux density, brightness temperature, stellar diameter) generally fall within the range of values
seen in previous measurements \citep[cf.][]{OG2015} there are hints of possible shifts in the relationships between these quantities. For example, in contrast to \cite{Lim1998}, who found $T_{\rm B}$ to continue to rise to $r\lsim$45~mas (the smallest projected radius probed by their 44~GHz measurement; see also \citealt{OG2015}) before declining at larger radii, all of our latest measurements (2019--2026) are consistent with the mean brightness temperature reaching its maximum at $r\sim$50--60~mas, and in 3 out of 4 epochs correspond to measurements made at $\nu\sim$22~GHz.   (Though it cannot be ruled out that the true $T_{\rm B}$ maximum may correspond to a slightly larger projected radius, as subsequent to the Great Dimming we have available only two measurements that sample $r>2.7R_{\star}$; see Figure~\ref{fig:tempsize}). 
This trend may be a manifestation of the change in the overall density structure of the 
chromosphere since the Great Dimming reported by \cite{Dupree2022}, although the sparsity of earlier radio measurements and the large  uncertainties in the individual $T_{\rm B}$ measurements make it impossible to determine conclusively that these changes were a direct result of the Great Dimming.

At frequencies above 44~GHz, only one prior $T_{\rm B}$ value has been published for Betelgeuse, namely the 2015 ALMA 338~GHz measurement from \cite{OG2017}. Our latest measurement in this band shows statistically significant decrease compared with 2015. Moreover, our 2021 and 2023 ALMA measurements in Bands 3, 4, 6, 7, and 8 appear to define an asymptotic decrease in mean brightness temperature between $27\lsim r \lsim 32$~mas, underscoring that the 2015 measurement is an outlier from the trend defined by more recent ALMA measurements. Given that the ALMA Band~3 and 4 measurements were separated from the Band 6, 7, and 8 measurements by approximately 2 years, this may reflect a persistent temperature decrease following the Great Dimming. However, follow-up measurements across additional epochs are need to establish whether the temperature profile interior to $r\lsim 2R_{\star}$ remains stable over a longer timeframe.

Another result that can be gleaned from Figure~\ref{fig:tempsize} is the confirmation of a temperature inversion or ``minimum'' between the photosphere and chromosphere of Betelgeuse. Such temperature minima are a common feature of atmospheric models of cool stars (e.g., \citealt{Ulmschneider1977}; \citealt{Basri1981}; \citealt{Hartmann1984}) and observational evidence for this behavior in the atmosphere of Betelgeuse was previously reported by \cite{Harper2001} (based on spatially unresolved data) and subsequently confirmed by \cite{OG2017} based on their spatially resolved ALMA observation at 338~GHz.  However, with our ALMA Band~8 (485~GHz) measurement, we are now able to probe the temperature structure even closer to the photosphere, and find that the mean brightness temperature continues to drop interior to $\sim$30~mas ($r\lsim 1.2R_{\star}$), reaching a value more than 1200~K cooler than the nominal photospheric temperature ($T_{\rm eff}\approx3476-3650$~K; see above).  Future spatially resolved observations with ALMA at even higher frequencies (Bands 9 and 10) should help to further constrain the precise location of the true temperature minimum.  We note also that results from infrared spectroscopy have long pointed to the existence of a cooler molecular atmosphere in Betelgeuse near $r\sim$1.2--1.5$R_{\star}$ know as the ``MOLsphere''. Temperatures for this layer derived by various authors range from $\sim$1500--2300~K (\citealt{Tsuji2000}; \citealt{Perrin2004}, \citeyear{Perrin2007}; \citealt{Ohnaka2004}; \citealt{Montarges2014}), thus it overlaps in both projected radius and temperatures measured by our latest ALMA observations (Figure~\ref{fig:tempsize}).  However, as pointed out by \cite{OG2017}, infrared measurements are insensitive to warmer gas that may be present at similar radii. On the other hand, spatially resolved radio observations provide a measure of the mean gas (electron) temperature. Thus our new measurements not only provide a more robust confirmation of the existence and location of the temperature minimum in the vicinity of the MOLsphere, but they imply that  gas much cooler than the photosphere truly dominates the volume density of this region. 

Finally, turning to larger radii, the mean brightness temperature as a function of radius remains poorly constrained for Betelgeuse beyond $r\gsim$80~mas owing to the large scatter is existing measurements. Probing this regime requires spatially resolved measurements at frequencies in the range $\sim$5--15~GHz, which are challenging owing to the decreasing source flux and the need for very long baselines to spatially resolve the star at these frequencies (e.g., \citealt{Richards2013}). Such measurements should become much more routine with future facilities including the Square Kilometer Array and the Next Generation Very Large Array (e.g., \citealt{Matthews2024}).

%
\begin{table}
\caption{New Radio Measurements of Betelgeuse}
\centering
\label{tab:measurements}
\begin{tabular}{ccccccccc}
\hline
Date  &  $\nu_{0}$ & Flux Density & $\theta_{\rm a}$ & $\theta_{\rm b}$ &  PA    & $e$ &D& $T_{\rm B}$\\ 
(UTC) &  (GHz) & (mJy)            & (mas)             & (mas)           &  (deg) &     &(au) & (K)\\ 
(1) &          (2) &            (3) &                (4) &     (5) &  (6) &   (7) &    (8)  & (9)\\
\hline
2023-09-09 & 22.0 & 11.4$\pm$1.3 (0.01) & 115.0$\pm$0.18 & 112.1$\pm$0.20 & 124.3$\pm$1.6 & 0.025$\pm$0.002 & 25.3&  3220$\pm$370 \\
2023-09-19 & 22.0 & 12.2$\pm$1.4 (0.01) & 114.5$\pm$0.14 & 111.9$\pm$0.15 & 107.4$\pm$1.2 & 0.023$\pm$0.002 &25.2&  3470$\pm$400\\
(Combined) & 22.0 &  11.8$\pm$1.3 (0.06) & 115.0$\pm$0.11 & 111.7$\pm$0.13 & 118.2$\pm$0.8 & 0.029$\pm$0.001 & 25.2 &3340$\pm$370 \\
2024-10-23 & 22.0 &  11.9$\pm$1.3 (0.01) & 120.7$\pm$0.14 & 108.1$\pm$0.17 & 96.4$\pm$0.3  &  0.104$\pm$0.002 & 25.4& 3320$\pm$360 \\
2026-04-04 & 22.0 &  12.3$\pm$1.4 (0.01) & 114.9$\pm$0.12 & 113.1$\pm$0.13 & 116.1$\pm$1.5 & 0.016$\pm$0.001 & 24.7 & 3460$\pm$390 \\
2023-09-09 & 44.0 & 32.9$\pm$3.3 (0.03) & 102.3$\pm$0.16 & 99.9$\pm$0.16 & 92.5$\pm$1.4 & 0.024$\pm$0.002 & 22.5&2930$\pm$290 \\
2023-09-20 & 44.0 & 32.2$\pm$3.2 (0.04) & 98.4$\pm$0.17 & 94.2$\pm$0.17 & 101.1$\pm$0.9 & 0.043$\pm$0.002 & 21.4& 3160$\pm$310\\
(Combined) & 44.0 & 32.5$\pm$3.3 (0.02) & 100.6$\pm$0.11 & 97.2$\pm$0.17 & 99.3$\pm$0.8 & 0.034$\pm$0.002 &22.0 & 3030$\pm$310\\
2024-10-23 & 44.0 & 35.3$\pm$3.5 (0.04) & 103.2$\pm$0.20 & 92.5$\pm$0.20 &  91.8$\pm$0.3 & 0.104$\pm$0.003 & 21.7 & 3370$\pm$330\\
2021-09-11 & 107.5 & 103.5$\pm$5.2 (0.02) & 77.3$\pm$0.03 & 72.0$\pm$0.02 & 146.5$\pm$0.1 & 0.068$\pm$0.001 & 16.6 & 2840$\pm$140\\
2021-09-12 & 135.6 & 135.4$\pm$6.8 (0.02) & 73.2$\pm$0.02 & 68.4$\pm$0.02 & 137.0$\pm$0.1 & 0.066$\pm$0.001 & 15.7 &  2590$\pm$130 \\
2023-08-03 & 223.7 & 277.0$\pm$13.8 (0.04) & 63.4$\pm$0.01 & 61.6$\pm$0.01 & 31.4$\pm$0.2 & 0.028$\pm$0.001 &13.9 & 2500$\pm$120\\
2023-08-27 & 223.7 & 271.7$\pm$13.5 (0.17) & 63.9$\pm$0.06 & 60.9$\pm$0.05 & 47.8$\pm$0.5 & 0.047$\pm$0.001 & 13.9 & 2460$\pm$120\\
(Combined) & 223.7 & 276.2$\pm$13.8 (0.03) & 63.5$\pm$0.01 & 61.4$\pm$0.01 & 38.9$\pm$0.1 & 0.033$\pm$0.001 & 13.9 & 2490$\pm$120\\ 
2023-08-03 & 337.8 & 500.4$\pm$35.0 (0.06) & 57.7$\pm$0.01 & 56.4$\pm$0.01 & 36.3$\pm$0.1 & 0.022$\pm$0.001 &12.7 & 2370$\pm$170 \\
2023-08-04 & 337.8 & 497.6$\pm$34.8 (0.08) & 58.1$\pm$0.01 & 56.5$\pm$0.01 & 37.3$\pm$0.2 & 0.028$\pm$0.001 &12.7 &  2340$\pm$160 \\
(Combined) & 337.8 & 499.4$\pm$35.0 (0.05) & 57.8$\pm$0.01 & 56.4$\pm$0.01 & 36.2$\pm$0.1 & 0.024$\pm$0.001 &12.7 &  2370$\pm$170\\
2023-08-04 & 485.2 & 862.0$\pm$86.2 (0.43) & 54.6$\pm$0.03 & 53.3$\pm$0.02 & 33.1$\pm$0.5 & 0.024$\pm$0.001 & 12.0 & 2220$\pm$220\\ 
\hline
\end{tabular}
\flushleft Flux density, angular dimensions, and position angles were derived by
fitting a uniform elliptical disk to the visibility data using AIPS task {\sc OMFIT}.  
Explanation of columns: (1) observing date; (2) mean observing frequency in GHz; (3) flux density
in mJy; (4) angular diameter of the major axis of the disk in
mas; (5)  angular diameter of the minor axis of the disk in mas; (6) PA of the major axis in
  degrees, measured east from north; (7) ellipticity, defined as
  $e=(\theta_{\rm a} - \theta_{\rm b})/(\theta_{\rm a})$; (8) mean stellar diameter in au, derived using the geometric mean angular
  diameter $\theta_{m}=\sqrt{\theta_{a}\theta_{b}}$ and adopting a stellar distance of 222~pc; (9) brightness
  temperature in Kelvin,  derived according to Equation~1 assuming a uniform elliptical
  disk model for the star. Quoted error bars for angular size and position angle measurements reflect only formal fitting uncertainties. 
  The errors bars for the flux density measurements also include the estimated contribution from the  uncertainty in the flux density scale in each band (see Section~\ref{unidisk}), with
  the contribution of the formal fit uncertainty indicated in parentheses. 

\end{table}
\subsection{Spectral Index}\protect\label{specindex}
Centimeter wavelength studies of Betelgeuse over the past few decades have consistently yielded a disk-averaged spectral index (i.e., a spectral index based on measurements of the integrated source flux) of $\alpha\approx1.33$, irrespective of flux variations in the star over time (\citealt{NewHjell1982}; \citealt{OG2015}; \citealt{Matthews2022}). Such a value of $\alpha$ is consistent with thermal emission originating predominantly from an optically thick extended atmosphere or chromosphere. 

\begin{figure*}
\begin{center}
	\includegraphics[width=15cm,angle=-180]{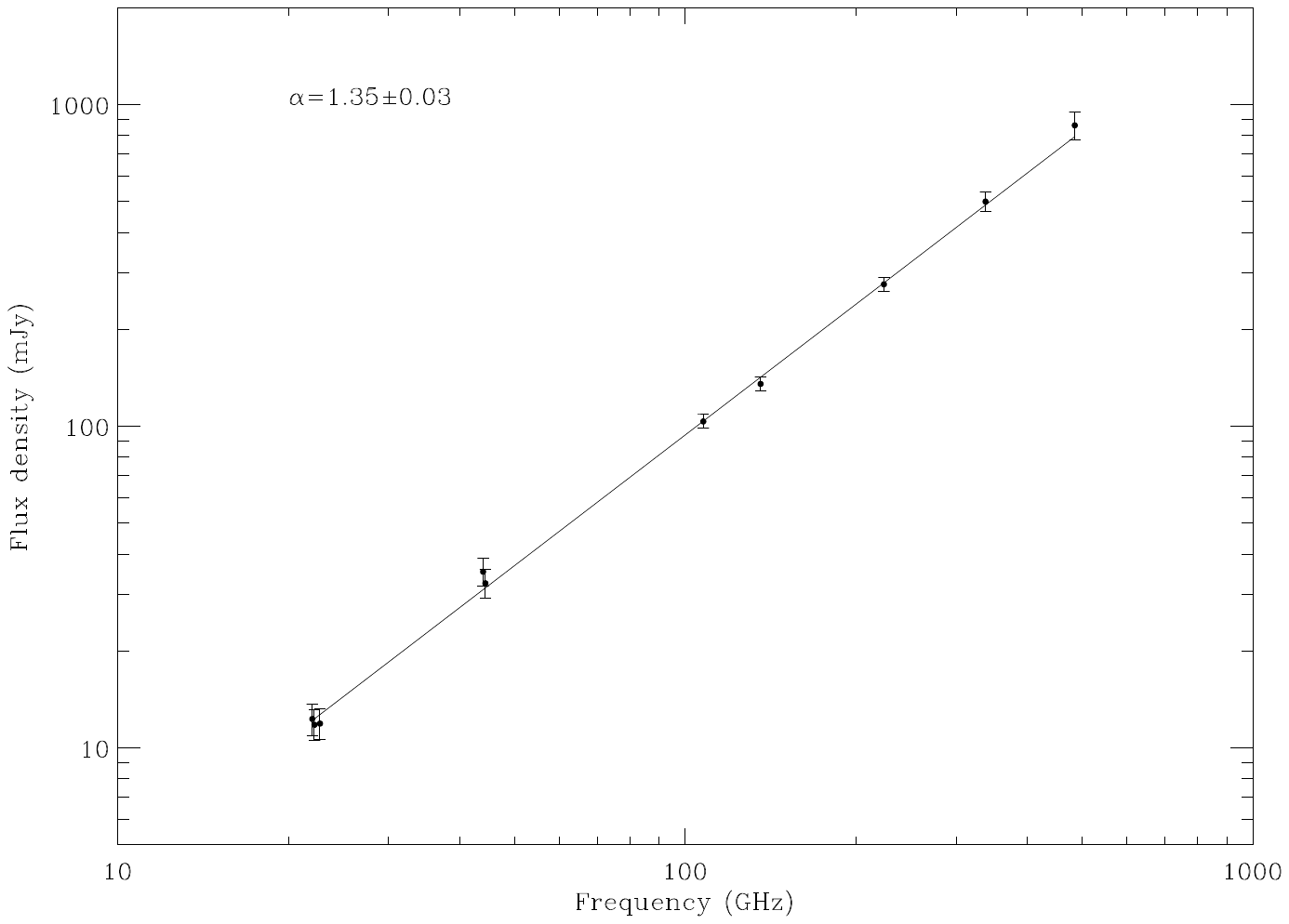}
    \caption{Radio spectrum of Betelgeuse, based on total flux density measurements derived from uniform elliptical disk fits to the visibility data (Table~\ref{tab:measurements}).  A power law fit to the data yields $\alpha=1.35\pm0.03$, consistent with past measurements at cm wavelengths (see text). Measurements obtained at identical frequencies were offset slightly along the $x$-axis for clarity.
    }
    \label{fig:specindex}
    \end{center}
\end{figure*}

Based on our latest measurements, which span more than two decades in frequency  (Table~\ref{tab:measurements}), we find $\alpha=1.35\pm0.03$, consistent with previous determinations. Also noteworthy is that we find no evidence for a statistically significant break or change in slope of the power law at the higher frequencies (Figure~\ref{fig:specindex}). Limiting the fit to only the three highest frequency ALMA bands we find $\alpha=1.40\pm0.06$, indistinguishable from the determination based on the entire frequency range. This implies that the higher frequency measurements do not appear to be significantly contaminated by circumstellar dust emission

\begin{figure*}
\begin{center}
	\includegraphics[width=15cm,angle=-180]{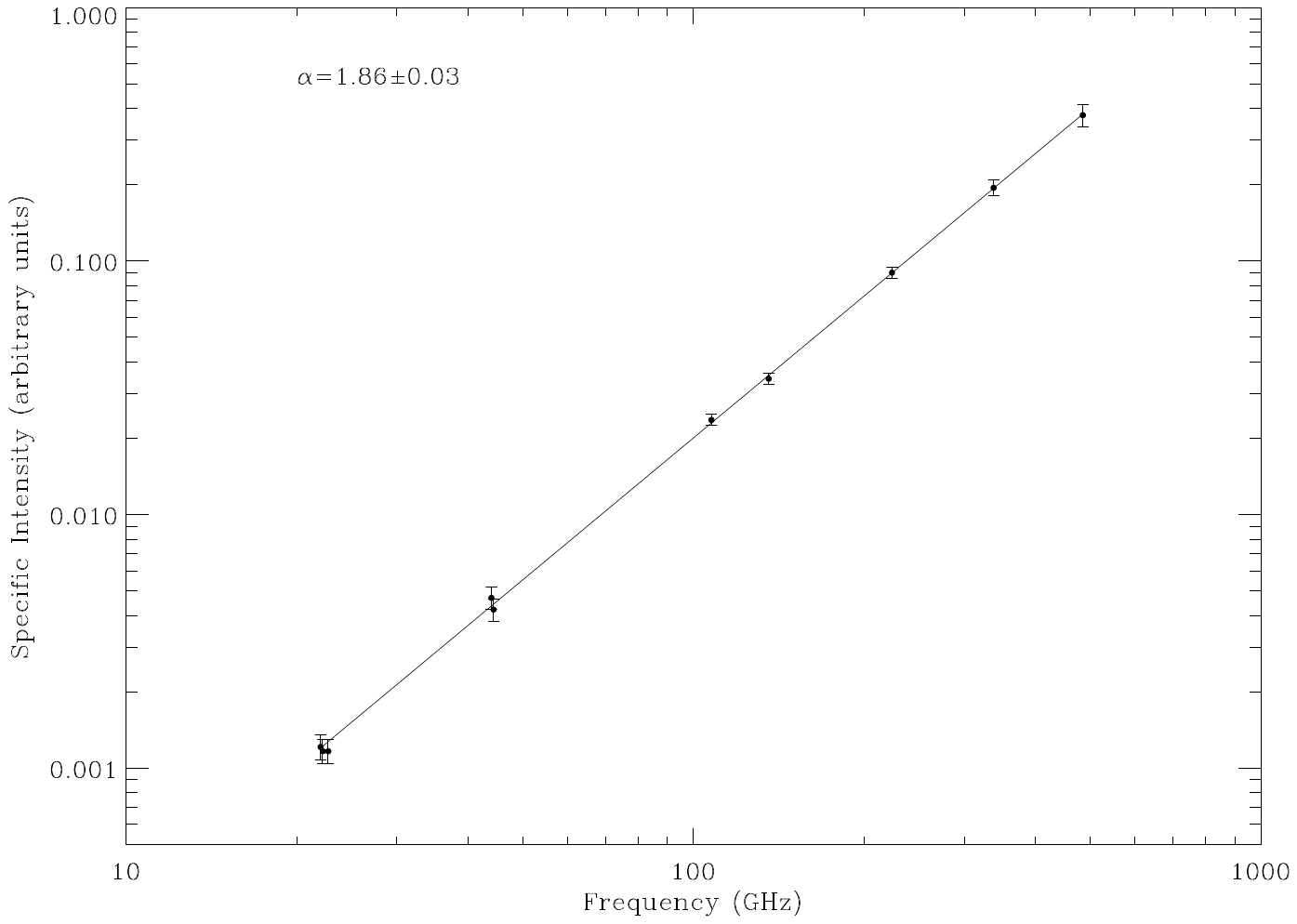}
    \caption{Specific intensity of Betelgeuse (in arbitrary units) as a function of frequency, based on parameters derived from uniform elliptical disk fits to the visibility data (see text for details). A power law fit to the data yields $\alpha_{\rm S}=1.86\pm0.03$. Measurements obtained at identical frequencies were offset slightly along the $x$-axis for clarity.
    }
    \label{fig:specific}
    \end{center}
\end{figure*}

Given that the star is spatially resolved in each of our current observations, we have also examined the spectral index derived from the {\em specific intensity} at each frequency, $\alpha_{\rm s}$. Here we define specific intensity (in arbitrary units) as: $S_{\nu}/(0.25\pi\theta_{a}\theta_{b}$), where $S_{\nu}$ is the global flux density in units of mJy and $\theta_{a}$ and $\theta_{b}$ and the major and minor axis diameters in units of mas, respectively (see Table~\ref{tab:measurements}). The result is  shown in Figure~\ref{fig:specific}. In this case we find $\alpha_{\rm s}=1.86\pm0.03$, close to the optically thick limit (where $\alpha=$2.0). This agrees well with the spectral index derived from spatially resolved ALMA images \citep{Dent2026}. As with the spectral index derived from the global source flux (Figure~\ref{fig:specindex}), no evidence is seen for a curvature or slope change in the power law over the frequency range covered by the current observations.



\subsection{Azimuthally Averaged Visibility Data\protect\label{azimuthal}}
\begin{figure*}
\begin{center}
  \includegraphics[width=16cm,angle=-180]{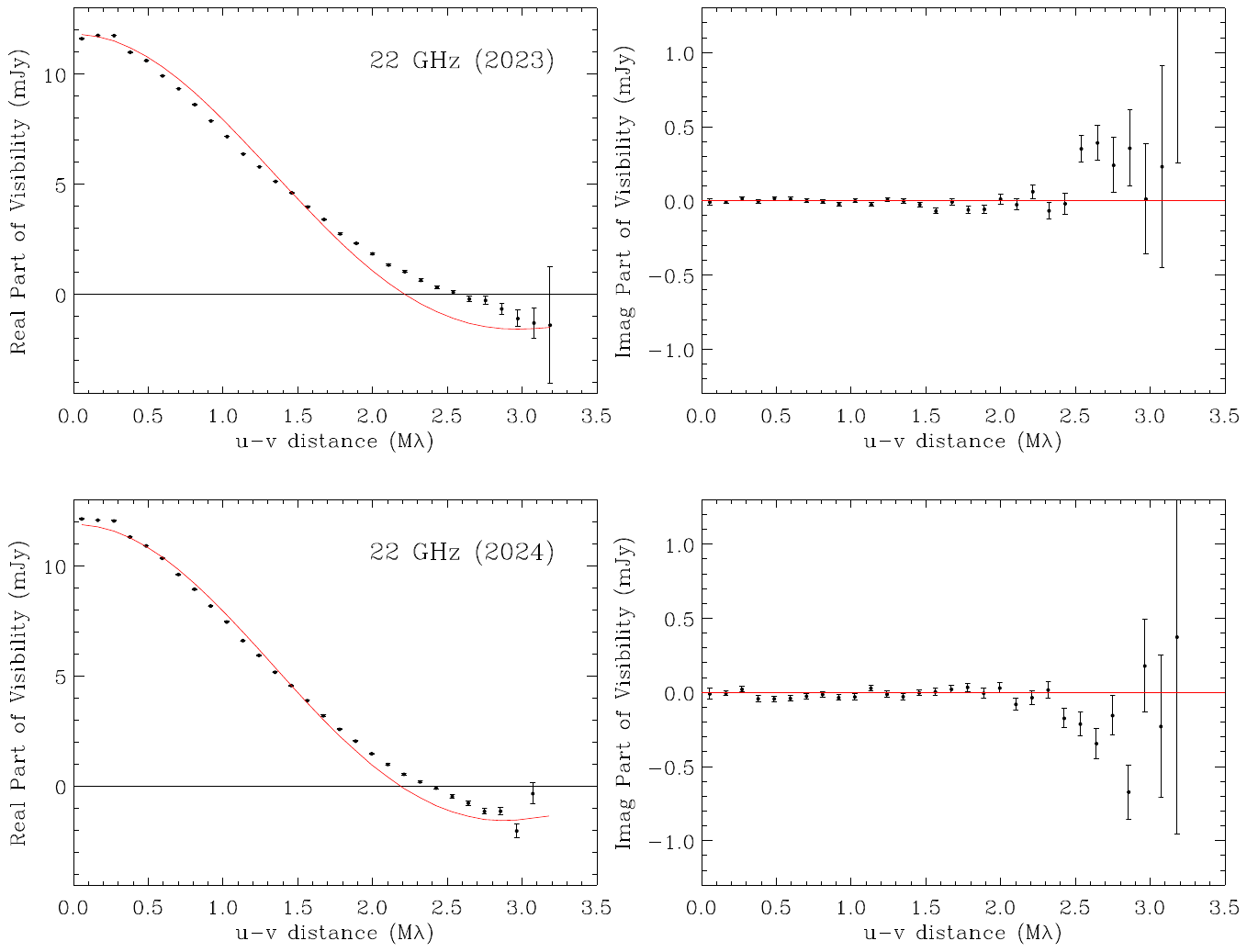} \vspace{-2.0cm}
  \includegraphics[width=16cm,angle=-180, trim=0cm 4cm 0cm 2cm]{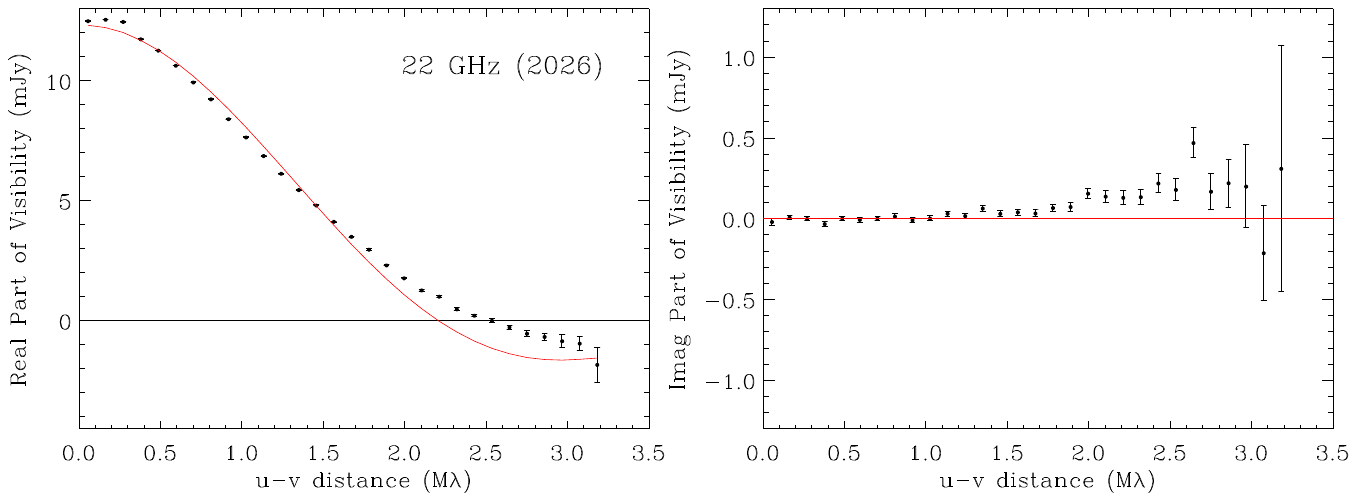}
     \vspace{-3.0cm}
    \caption{Azimuthally averaged visibility data based on 22~GHz VLA observations from 2023 (top row), 2024 (middle row), and 2026 (bottom row). For 2023 the combined data from both observing epochs 
    were used. The real part of the visibilities (in mJy) is plotted as a function of $u$-$v$ distance (in M$\lambda$) in the left panels, while the right panels show the imaginary part of the visibilities. The best-fitting uniform elliptical disk models from Table~\ref{tab:measurements} are overplotted with red lines. 
    }
    \label{fig:Kvisplots}
    \end{center}
\end{figure*}

\begin{figure*}
\begin{center}
  \includegraphics[width=14cm,angle=-180]{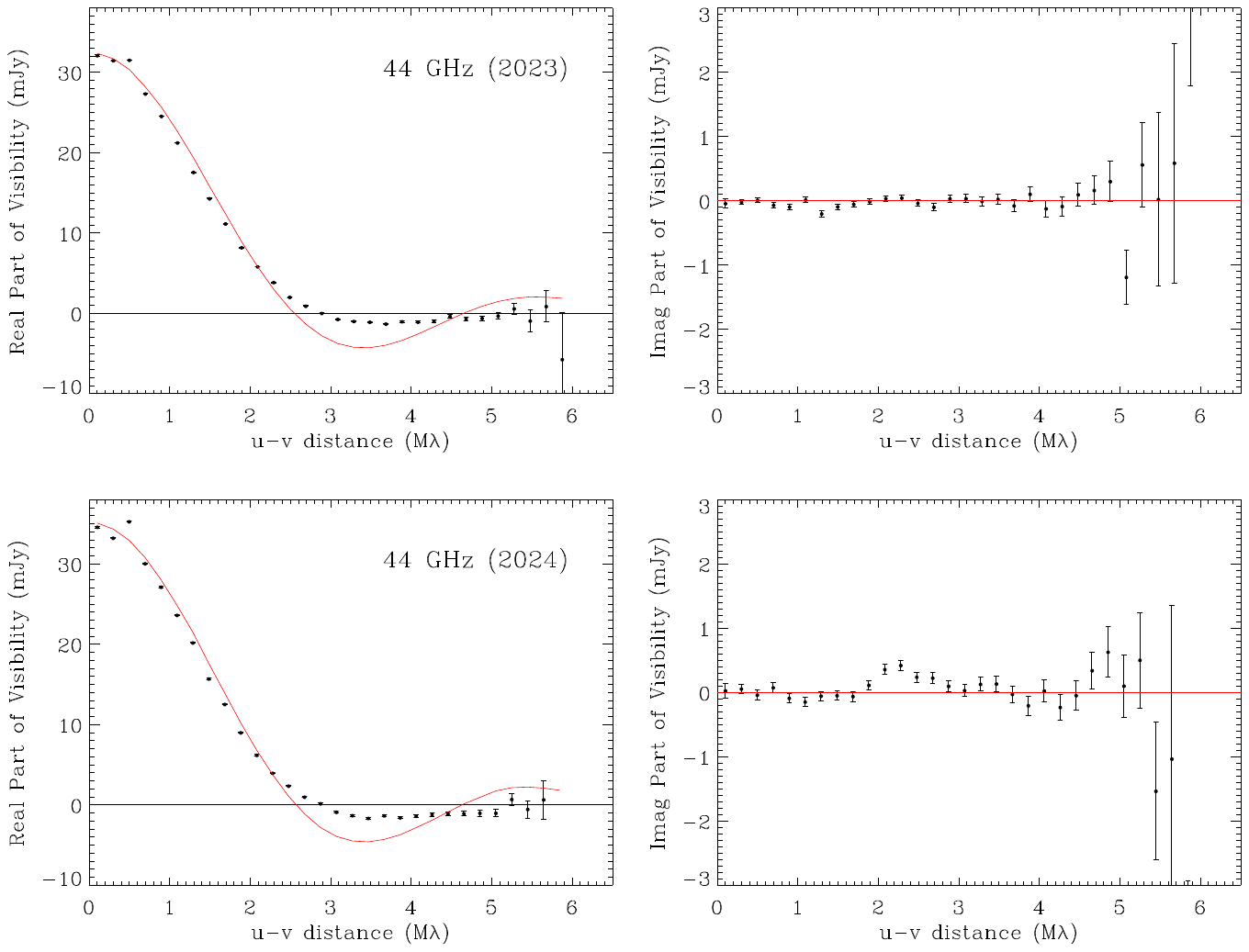}
    \caption{As in Figure~\ref{fig:Kvisplots}, but for the VLA 44~GHz data.
    }
    \label{fig:Qvisplots}
    \end{center}
\end{figure*}

For each observed frequency band in the present study we have derived azimuthal averages of the real and imaginary components of the visibility data. The results are presented in Figures~\ref{fig:Kvisplots}--\ref{fig:B8visplots}.  The real component of the visibilities provides information on the angular extent and structure of the emission, while the imaginary components provide insights into the level of symmetry of the emission.  
For reference, 
on each of the panels in Figures~\ref{fig:Kvisplots}--\ref{fig:B8visplots} we overplot the best-fitting uniform elliptical disk model from Table~\ref{tab:measurements}. 

We see that for the various ALMA bands, to first order, the uniform elliptical disk model provides a reasonable fit to the real part of azimuthally-averaged visibilities over the range of angular scales covered by our observations. However, at cm wavelengths (22 and 44 GHz frequencies), it is clear that the elliptical disk model systematically underestimates the stellar emission on angular scales probed by the longest baselines [$\gsim$2--3~megalambda (M$\lambda$)]. This was also seen in the 2019 observations of Betelgeuse from \cite{Matthews2022}. At frequencies $\nu\lsim$20~GHz, there is evidence that the radio emission from red supergiants becomes increasingly dominated by the stellar wind rather than the chromosphere (\citealt{Alt1979}; \citealt{OG2020}), thus one explanation is that our lowest observing frequencies we are starting to see a contribution from the stellar wind and/or wind acceleration region at radii beyond a few $R_{\star}$ (e.g., Figure~3 of \citealt{OG2020}). The density distribution of the emission on these scales may also be impacted by the wake of the orbiting companion (see Section~\ref{companion}).

Unlike radio interferometric images, visibility data are not affected by the choice of the adopted weighting scheme and are not subject to possible deconvolution artifacts,  hence they are especially useful for assessing general properties of the radio emission and for providing a benchmark for testing the predictions of atmospheric models (e.g., \citealt{Harper2001}). A companion study by \cite{Dent2026} utilizes the new visibility data presented here as input for a new spatially resolved semi-empirical thermodynamic model of the extended atmosphere and wind of Betelgeuse. The broad frequency coverage of our current study helps to enable characterization of the full range emitting material along the line-of-sight to the deeper atmospheric layers probed by our highest frequency ALMA bands. This, coupled with 
the improved signal-to-noise ratio of the VLA visibility data compared with earlier work,
has enabled significant improvements over previous semi-empirical models \citep[cf.][]{Harper2001}.

\begin{figure*}
\begin{center}
  \includegraphics[width=14cm,angle=-180]{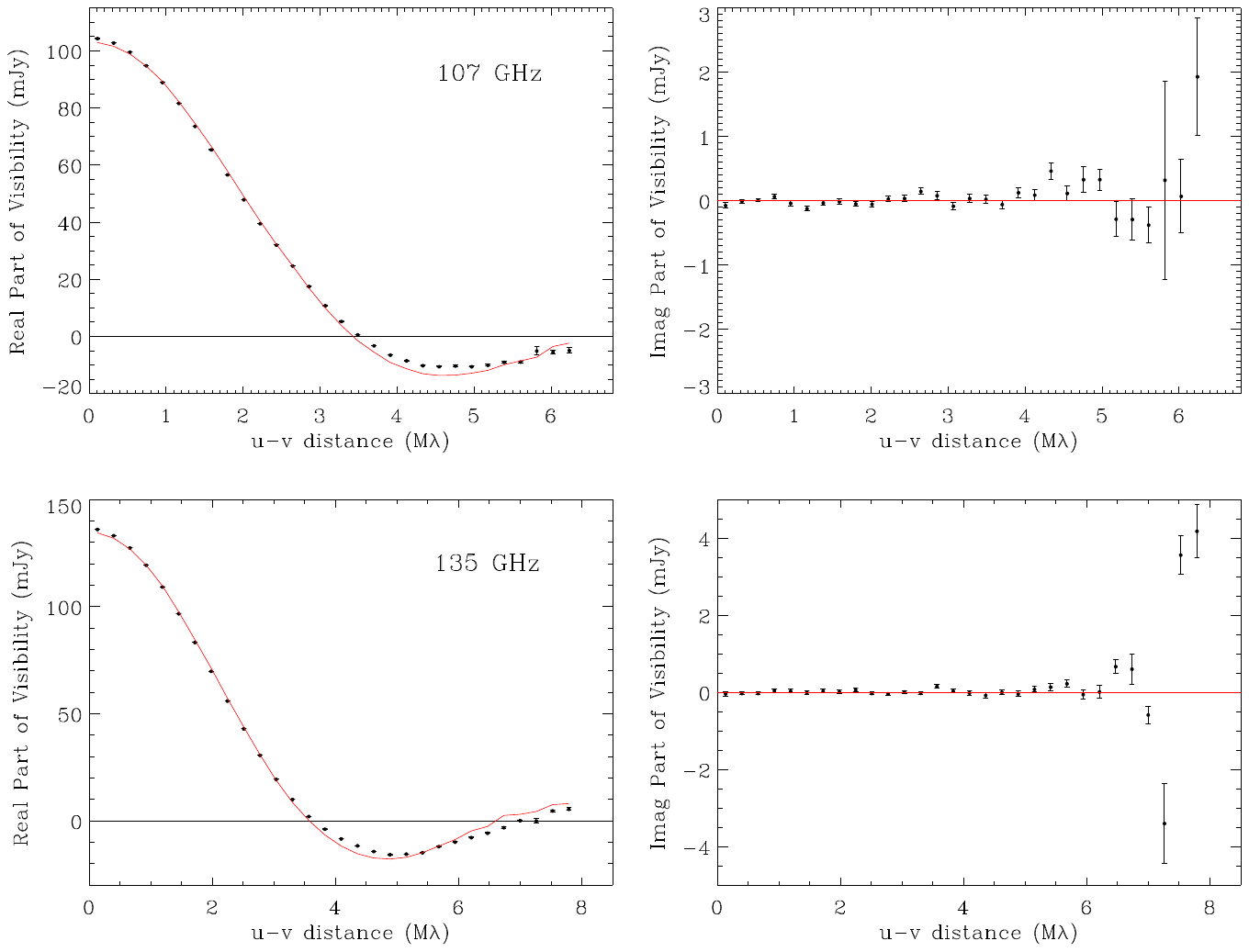}
    \caption{As in Figure~\ref{fig:Kvisplots}, but for the ALMA 107~GHz and 135~GHz data.
    }
    \label{fig:B3-4visplots}
    \end{center}
\end{figure*}
\begin{figure*}
\begin{center}
  \includegraphics[width=14cm,angle=-180]{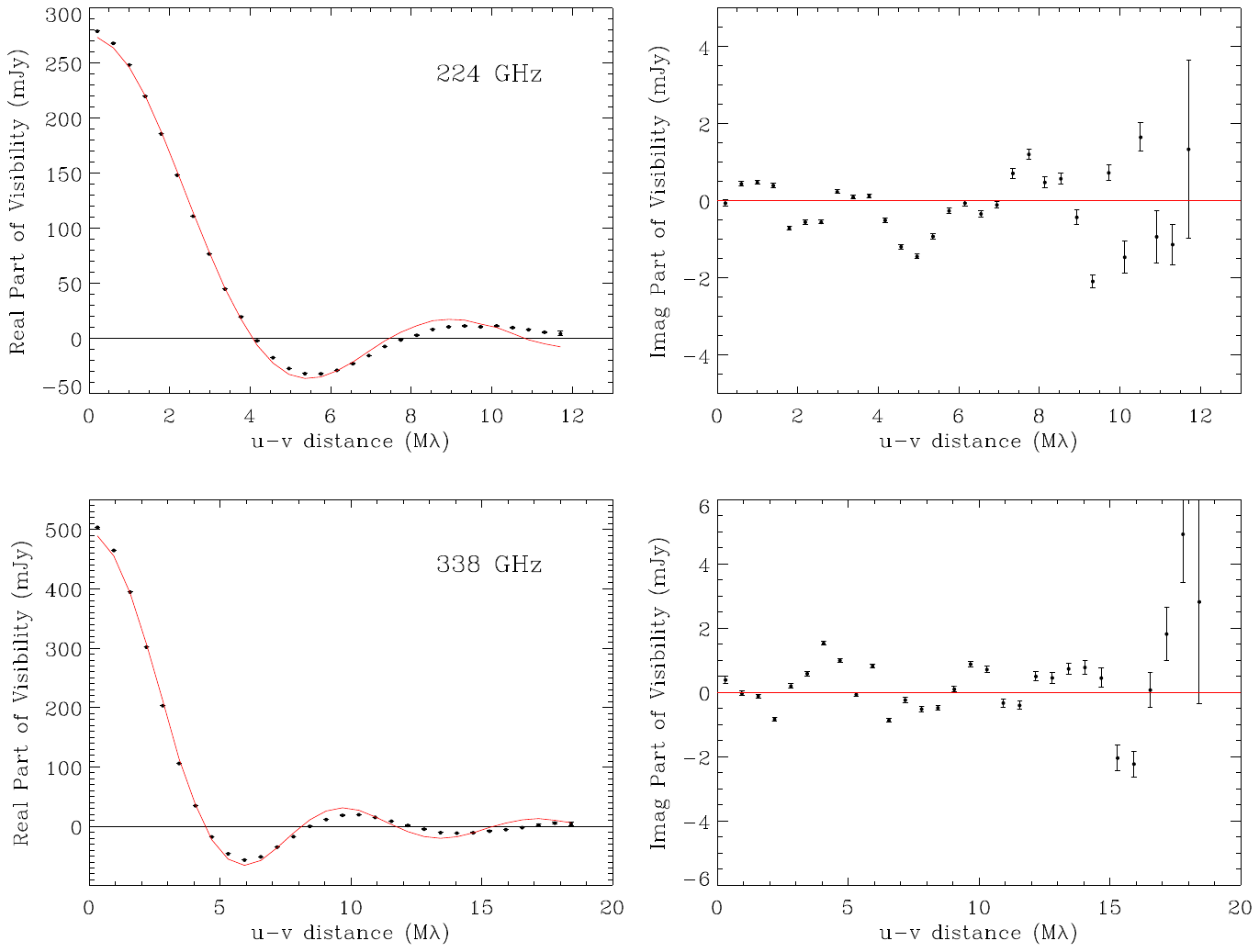}
    \caption{As in Figure~\ref{fig:Kvisplots}, but for the ALMA 224~GHz and 338~GHz data.
    }
    \label{fig:B3-4visplots}
    \end{center}
\end{figure*}
\begin{figure*}
\begin{center}
  \includegraphics[width=14cm,angle=-180]{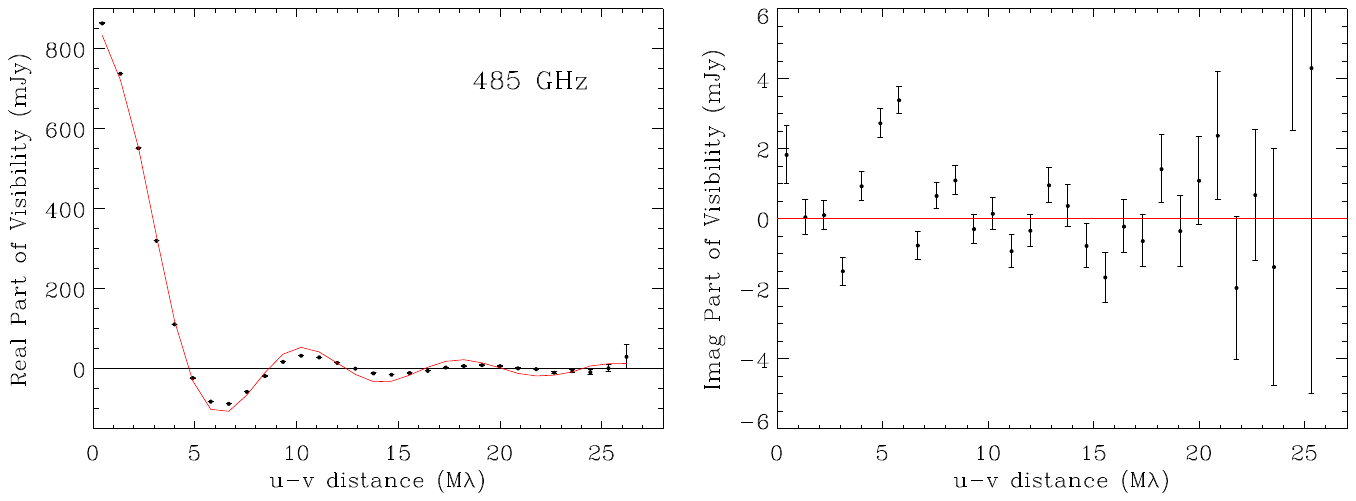}
    \vspace{-5.0cm}
    \caption{As in Figure~\ref{fig:Kvisplots}, but for the ALMA 485~GHz data.
    }
    \label{fig:B8visplots}
    \end{center}
\end{figure*}

\section{A Search for Radio Wavelength Signatures of a Companion to Betelgeuse\protect\label{companion}}
Two independent studies recently raised the suggestion that the long secondary period of Betelgeuse ($\sim$2100~d) is caused by a binary companion orbiting within the extended atmosphere of the star, with a semi-major axis $a\approx$50~mas, or equivalently, $\approx2.3R_{\star}$ (\citealt{Goldberg2024}; \citealt{Mac2025}). Subsequently, using optical speckle imaging, 
\cite{Howell2025} reported a tentative direct detection of this companion, which they named ``Siwarha'' (now an official International Astronomical Union designation)\footnote{\url{https://www.iau.org/IAU/IAU/News/Ann2026/New-Star-Names-2026.aspx}}, at an angular separation of 52~mas and PA=115$^{\circ}$ near the time
when it was predicted to be at maximum elongation (orbital phase 0.25).\footnote{Another maximum elongation occurs at phase 0.75, 
when the companion is on the opposite side of the star.}

Evidence for the companion was further strengthened by the multi-epoch
spectroscopic study of \cite{Dupree2026}. These authors identified variability in the optical circumstellar \MnI\ line that is consistent with the long secondary period of Betelgeuse, and
adopting the orbital model of \cite{Mac2025}, they found that the circumstellar 
absorption increases immediately after the expected transit of the companion across the disk
of Betelgeuse (i.e., phase 0.0). The level of absorption then reaches a maximum close to the companion eclipse phase
($\sim$0.5) before decreasing, returning to its initial state after $\sim$2100 days. The \MnI\ lines also
show variations in expansion velocity relative to the photosphere that are consistent with the
long secondary period. \cite{Dupree2026} additionally presented multi-epoch ultraviolet
spectra of several chromospheric emission lines obtained with the {\it Hubble Space Telescope}
between 2019 and 2025. These ultraviolet lines display asymmetry changes and velocity variations
that are strongly correlated with the predicted orbital phase of the companion and
the authors interpret these phase-dependent variations as resulting from velocity and
density variations induced by an expanding wake trailing the orbiting secondary.

Perhaps the most compelling evidence for the companion now comes from the recent study of \cite{Montarges2026}. These authors used
direct imaging with the Very Large Telescope SPHERE-ZIMPOL adaptive optics system to obtain a 6$\sigma$ detection of the companion in the optical at its predicted
position. 

\subsection{Prospects for Direct Radio Detection of the Companion}
The exact nature of Siwarha is still controversial (see \citealt{Mac2025}; \citealt{Howell2025}; \citealt{Montarges2026}), but regardless of its exact mass and spectral type, the thermal radio emission from the companion itself is expected to be too weak to detect in our current observing bands. For example, assuming a companion radius of 2$R_{\odot}$ and a temperature $T_{\rm eff}$=12,000~K \citep{Montarges2026}, the thermal blackbody flux at 44~GHz would be only $S_{\nu}\approx0.1~\mu$Jy, more than an order of magnitude too weak to be detected within a plausible VLA integration time. While blackbody flux increases with frequency, the predicted flux density in ALMA Bands~7 and 8 ($\nu\approx$338~GHz and $\nu\approx$485~GHz) is only $\sim$5.4$\mu$Jy and $\sim 11\mu$Jy, respectively, below the rms noise in our images in both bands (see Table~\ref{tab:newimages}).
Unsurprisingly, we do not find any evidence of a direct detection of a radio counterpart to the companion in our  present data. 

Further confounding any prospect for direct radio detection in our present data is the predicted location of Siwarha during our recent ALMA observations. Based on the emphemeris of \cite{Mac2025} we provide in Table~\ref{tab:siwarha} an estimated PA and projected separation of the  companion relative to Betelgeuse for each of the dates on which new spatially resolved radio observations were obtained. We see that during our Band~6, 7, and 8 observations, the orbital phase of the companion was $\sim$1.08, with an angular offset from the center of Betelgeuse of only $\sim$23~mas (to the southeast), which is comparable to the stellar radius at this frequency (Table~\ref{tab:measurements}),

We note that if Siwarha is a young active/flaring star, the possible existence of time-variable nonthermal radio emission cannot yet be excluded. However, our current observing bands are not optimal for detecting such emission, which is more commonly seen at longer wavelengths ($\nu\sim$5--10~GHz; e.g., \citealt{Forbrich2016}).

\subsection{Possible Effects of the Companion on the Radio Emission of Betelgeuse}
\begin{figure*}
\begin{center}
  \includegraphics[width=0.4\textwidth,angle=0]{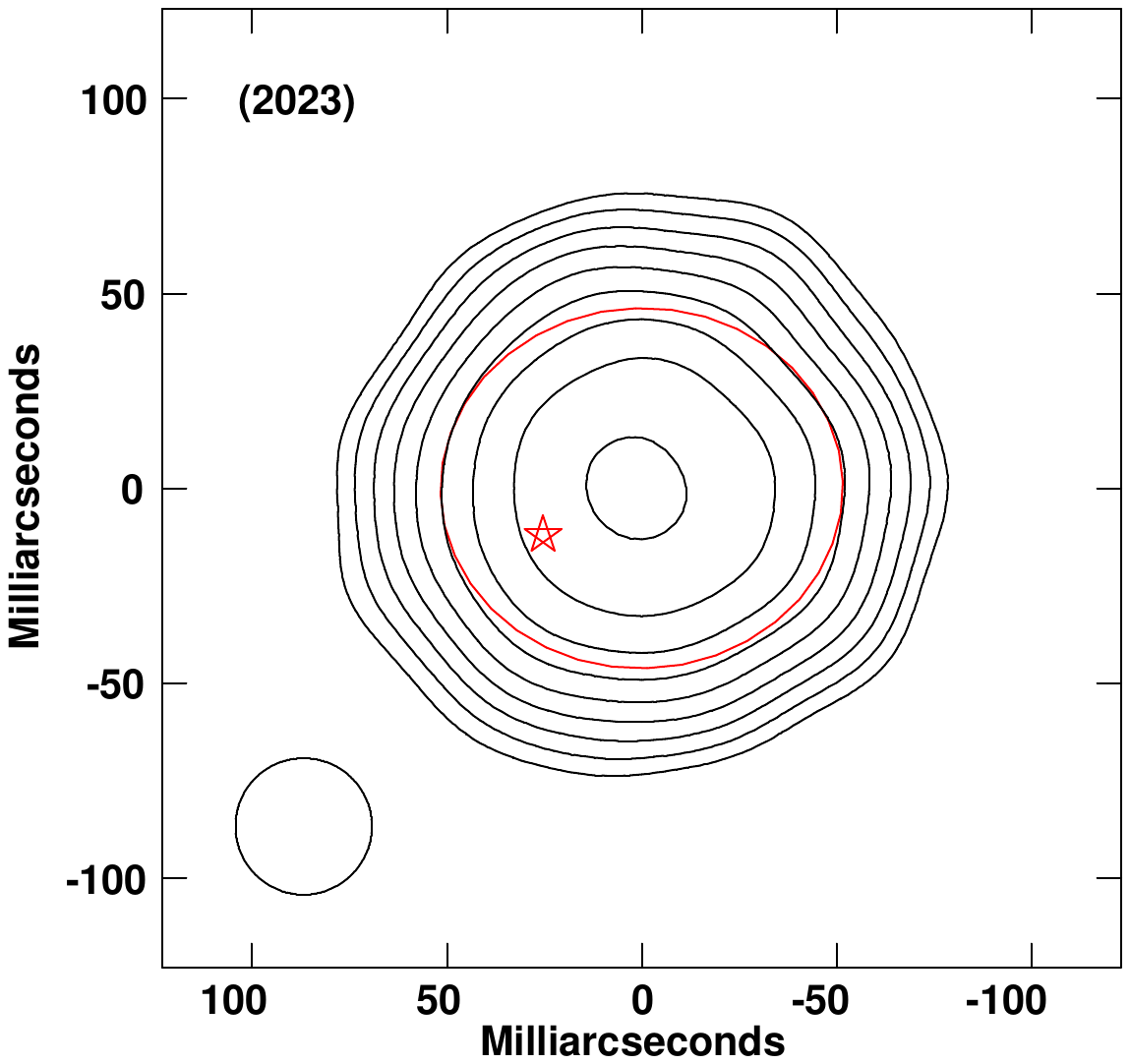}\hspace{-2.6cm}
	\includegraphics[width=0.4\textwidth,angle=0]{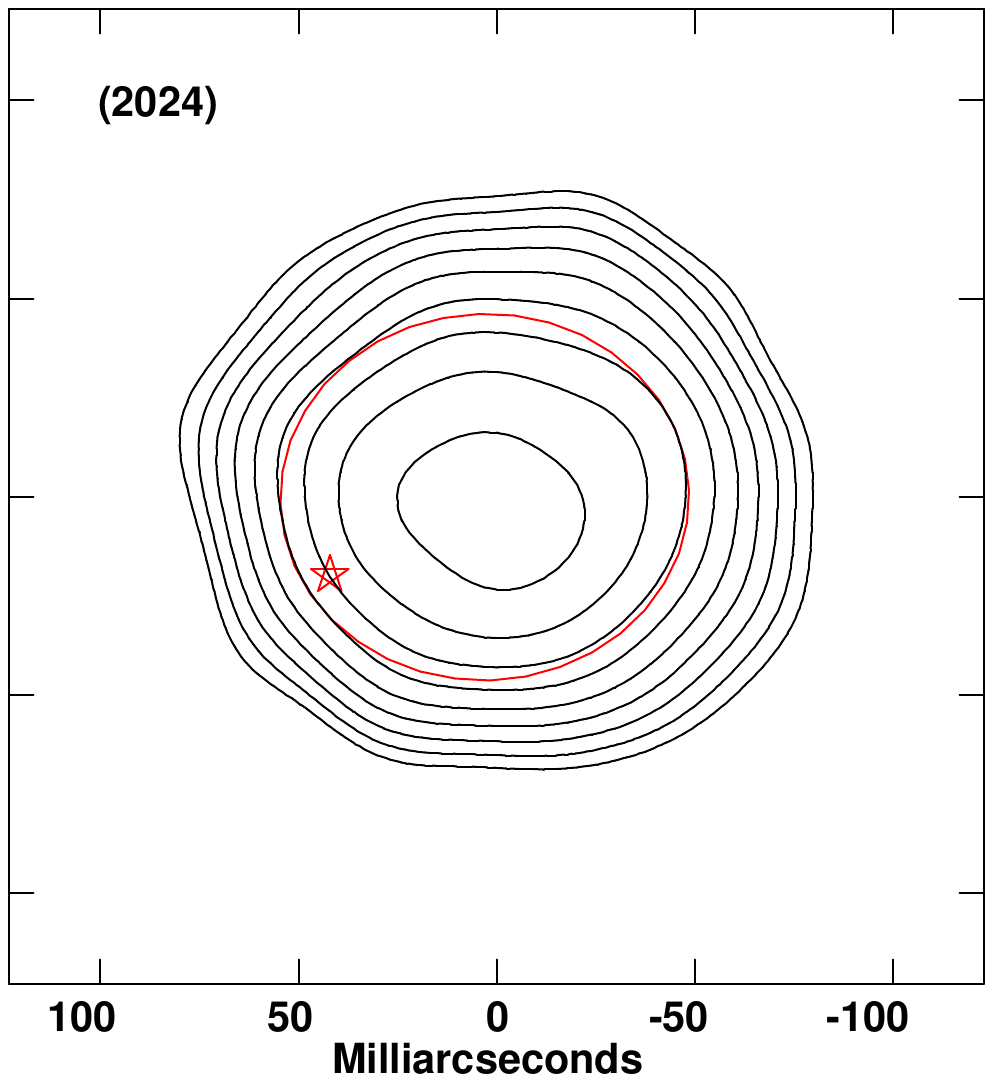}
    \caption{VLA contour maps of Betelgeuse at 44~GHz from 2023 (left) and 2024 (right). The 2023 image uses the combined data from the two 2023 observing epochs (see Table~\ref{tab:newobs}). Both images were made using ${\cal R}$=$-1$ weighting and a circular restoring beam with FWHM 35~mas (indicated in the lower left panel). The rms noise levels are 22.7~$\mu$Jy beam$^{-1}$ (37.9~$\mu$Jy beam$^{-1}$) for 2023 (2024). Contours in both images are ($-$10 [absent], 10, 14.1, 20, ...283, 400)$\times$37.9$~\mu$Jy beam$^{-1}$.  The overplotted red ellipses indicate the dimensions of the best-fitting uniform elliptical disk model from Table~\ref{tab:measurements}.  The red star symbols indicate the estimated projected position of the companion in the plane of the sky during the epoch of observation based on the ephemeris of \cite{Mac2025}. In 2023 the companion was slightly past transit, while in 2024 it was close to maximum elongation (see Table~\ref{tab:siwarha}).   }
    \label{fig:companionlocation}
    \end{center}
\end{figure*}

Even if the radio emission from Siwarha is intrinsically too faint to detect directly in our VLA and ALMA observations, 
the companion's orbital properties raise the interesting prospect of detecting its signatures {\it indirectly}. 
Indeed, the inferred semi-major axis of the companion's orbit ($\sim2.3R_{\star}$; \citealt{Mac2025}) implies that it orbits {\it within} the radio-emitting regions of the stellar chromosphere of Betelgeuse. To illustrate this, in Figure~\ref{fig:companionlocation} we overplot on our 2023 and 2024 VLA 44~GHz images the predicted projected position of the companion at the corresponding date.  Such a close-in orbit may plausibly impact the properties of the radio atmosphere of Betelgeuse itself, e.g., by inducing distortions or asymmetries as a result of  ionization changes and/or dynamical perturbations.  

If Siwarha is a B-type star with $T_{\rm eff}\approx$12,000~K, it may plausible impact the ionization properties of the atmospheric material through which it is orbiting. We have therefore examined our recent images for signatures of asymmetries or brightness variations in the radio emission near the predicted position of the companion. However, 
do not see any such signatures that can be unambiguously linked with the companion phase. 
However, owing to the modest spatial resolution of our VLA and ALMA Band 3 and 4 images and the unfavorable orbital phase of the higher frequency ALMA images.

\begin{figure*}
\begin{center}
	\includegraphics[width=15cm,angle=0]{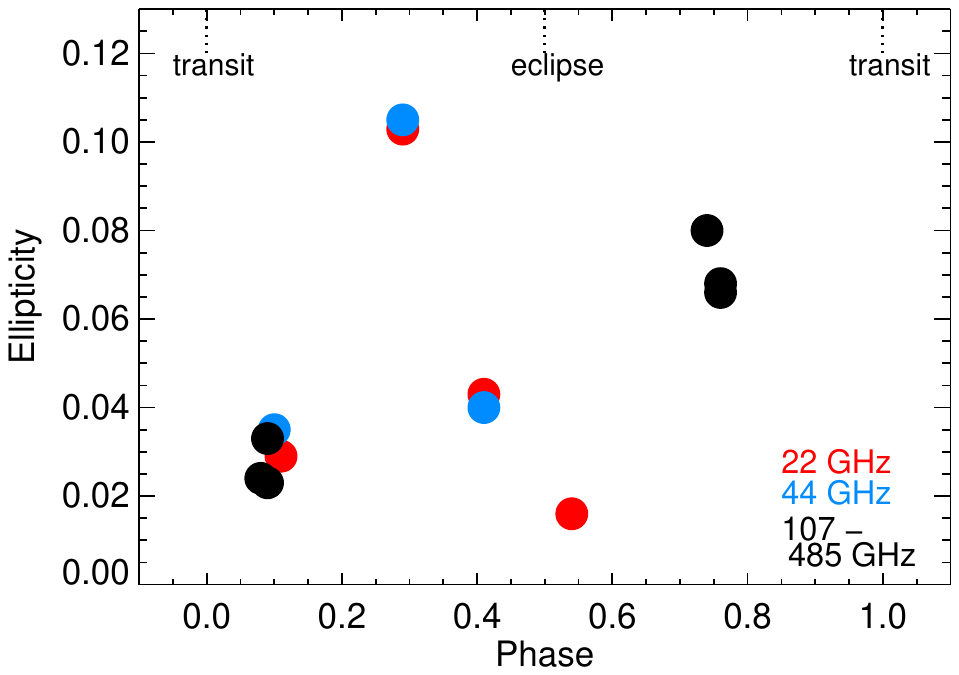}
     \vspace{-8.0cm}
    \caption{Ellipticity of Betelgeuse, as measured from multi-epoch ALMA and VLA measurements, plotted as a
function of the orbital phase of the suspected close-in companion (see Table~\ref{tab:siwarha}).  The colors of the points correspond
to different radio frequencies. Error bars on the ellipticity are smaller than the plotted symbols (see Table~\ref{tab:measurements}).
    }
    \label{fig:ellipticity}
    \end{center}
\end{figure*}

Based on the information from Table~\ref{tab:siwarha}, along with the stellar parameters from Table~\ref{tab:measurements}, we plot in Figure~\ref{fig:ellipticity} the measured ellipticity of Betelgeuse as a function of orbital phase of the companion. In addition to the results from the present study, we have included three additional data points from the spatially resolved observations of \cite{OG2017} (at 337~GHz) and \cite{Matthews2022} (at 22 and 44~GHz).
While as noted above, the ellipticity of the star is generally small in all observed bands and epochs, we nonetheless see evidence of a systematic increase in the ellipticity of the star close to times of maximum elongation of the companion (phases 0.25 and 0.75) and an ellipticity that approaches 0 during times of transit and eclipse (phases 0 and 0.5, respectively). These results suggest that the companion may indeed be having important impacts on the density, structure, temperature, and other properties of the outer atmosphere of this red supergiant. 

One puzzle concerning the above result is that while the overall ellipticity of Betelgeuse appears to correlate with orbital phase of the companion over a range in observing frequency, the inferred PA of the best-fitting ellipse shows significant differences among frequency bands. The highest frequency ALMA bands (6, 7, and 8) yield PA values in the range 31 to 47 deg, comparable to the value of PA=53~deg reported by \cite{OG2017} in an earlier Band~7 measurement.  In contrast, ALMA Bands~3 and 4 both show PA$\sim$140~deg. Meanwhile the VLA bands show PA values ranging from $\sim$90 to 120~deg (see Table~\ref{tab:measurements}), comparable to values reported at these frequencies by \cite{OG2015} and \cite{Matthews2022}. Because the different frequencies trace different depths  in the atmosphere, one possible explanation is that the different bands are tracing different parts of a three-dimensional density perturbation, such as the wake trailing the orbiting companion proposed by \cite{Dupree2026}.
Additional epochs of measurements spanning a wide range of both observing frequency and orbital phase of the companion will be  needed to establish whether these frequency-dependent PA values remain consist over both time and orbital phase.

%
\begin{table}
\caption{Predicted Location of Betelgeuse's Companion During Recent Radio Observations\protect\label{tab:siwarha}}
\centering
\label{tab:siwarha}
\begin{tabular}{ccccc}
\hline
Date & JD & Orbital Phase & $\Theta$ (mas) & Offset Direction\\ 
(1) &   (2) &  (3) &  (4) &  (5) \\
\hline
2015-11-09  &   2457335.5    &   0.74   &  $-$48.2  & NW \\
2019-08-02   &  2458697.5     &  1.39   & 25.9  & SE \\
2021-09-11/12 & 2459469.0      & 1.75   & $-$48.3 & NW \\
2023-08-03/04  & 2460160.0      & 2.08    & 23.3 & SE \\
2023-08-27    & 2460183.5    &   2.09   & 25.9 & SE \\
2023-09-09    & 2460196.5     &  2.10    &   28.4 & SE \\
2023-09-19/20 &  2460207.0     &  2.10  &   28.4 & SE \\
2024-10-23    & 2460606.5      & 2.29   & 46.8  & SE \\
2026-04-04    & 2461135.0      & 2.54   & 12.0 & NW \\
\hline
\end{tabular}
\flushleft{Predicted location of Betelgeuse's companion Siwarha, based on the ephemeris of \cite{Mac2025}, for dates corresponding to our current radio observations (cf. Table~1), along with the spatially resolved observations from \cite{OG2017} and \cite{Matthews2022}.
Explanation of columns: (1) calendar date; (2) Julian date; (3) orbital phase; (4)  angular separation from the center of Betelgeuse, along PA$\approx +115^{\circ}$ (in mas); (5) direction of offset of the companion relative to Betelgeuse in the plane of the sky. }

\end{table}

\section{Conclusions}
We have presented new spatially resolved imaging observations of Betelgeuse in seven different wavebands spanning a frequency range of 22$-$485~GHz ($\lambda$0.63~mm to $\lambda$1.4~cm).
The observations were obtained between 2021 and 2026, subsequent to Betelgeuse's historic Great Dimming and enable characterization of the properties of the extended atmosphere of the star between projected radii of $r\sim1.2-2.6R_{\star}$.

At 22 and 44~GHz, respectively, we find that the disk-averaged brightness temperatures of Betelgeuse have returned to nominal values following the historic lows observed in 2019 August \citep{Matthews2022}. The peak brightness temperatures seen in our latest measurements ($T_{\rm B}\approx$3500~K, corresponding to measurements made at 22~GHz) are now observed to reach values comparable to the photospheric effective temperature of Betelgeuse.

Based on the total flux density measured in each observed band we derive a spectral index
 $\alpha=1.35\pm0.03$, consistent with past cm wavelength measurements over multiple epochs. The data, which span more than two decades in frequency, are well fit with a single power law. The specific intensity  as a function of frequency is also well fit with a single power law with 
 $\alpha=1.86\pm0.03$, consistent with optically thick emission.

Given the optically thick nature of the radio emission over the frequency range covered by our observations, higher frequencies probe increasingly deeper layers in the star. Examining the mean (disk-averaged) brightness temperature, $T_{\rm B}$, as a function of frequency, our new measurements indicate that since at least 2019, $T_{\rm B}$ consistently peaks at $r\gsim2.7R_{\star}$, compared with $r\sim2.5R_{\star}$ based on studies prior to 2016. This is consistent with a possible recent evolution in the density structure of Betelgeuse's extended atmosphere.  We also confirm previous indications of the existence of a temperature inversion between the photosphere and the chromosphere, with 
 a roughly monotonic decrease in $T_{\rm B}$ from $\approx3500$~K at $r\sim2.7R_{\star}$ to $\approx$2220~K at $\sim1.2R_{\star}$. This minimum temperature is comparable to that previously measured in the infrared for the molecular layer known as the MOLsphere at comparable radii.  Spatially resolved radio measurements at higher frequencies (e.g., ALMA Bands 9 and 10) are needed to probe still smaller projected radii and better constrain the true location of the temperature minimum.

Using our latest measurements and other recent measurements from the literature we have found evidence that the ellipticity of Betelgeuse as measured in various radio bands is correlated with the orbital phase of its recently identified close-in companion, Siwarha, whose semi-major axis ($\sim2.3R_{\star}$) places its orbit within the radio-emitting atmosphere of Betelgeuse. This suggests that the companion is shaping the structure and density of the atmosphere of this red supergiant proximate to the region where the stellar wind is launched.
 
\begin{acknowledgments}
The authors thank G. Harper and E. O'Gorman for valuable discussions.
LDM and AKD are supported by awards HST-GO-16655.008-A and HST-GO-17522.007-A from the Space Telescope Science Institute, which is operated by the Association of Universities for Research in Astronomy, Incorporated, under NASA contract NAS5-26555. LDM was additionally supported by award AST-2107681 from the National Science Foundation. 
This paper makes use of the following data sets: ST2177 (VLA); SH31627 (VLA); 26A-591 (VLA);
ADS/JAO.ALMA\#2019.1.01098.S (ALMA); ADS/JAO.2022.A.00026.S (ALMA).  
ALMA is a partnership of ESO (representing its member states), NSF (USA) and NINS (Japan), together with NRC
(Canada), MOST and ASIAA (Taiwan), and KASI (Republic of Korea), in cooperation with the Republic of Chile. The Joint ALMA Observatory is operated by ESO, AUI/NRAO and NAOJ.
 
\end{acknowledgments}

%

\vspace{5mm}
\facilities{VLA, ALMA}


\software{AIPS \citep{Greisen2003},  
          CASA \citep{CASA2022} }




\bibliography{my_biblio}{}
\bibliographystyle{aasjournalv7}



\end{document}